\documentclass[a4paper,USenglish]{lipics-v2018}
\usepackage{microtype}
\usepackage[boxed,vlined,linesnumbered]{algorithm2e}
\usepackage{amsmath}
\usepackage{amssymb}
\usepackage{comment}
\usepackage{csquotes}
\usepackage{enumitem}
\usepackage{etex}
\usepackage{mathdots}
\usepackage{mathtools}
\usepackage{pgfplots}
\usepackage{pgfplotstable}
\usepackage{subcaption}

\title{On Undetected Redundancy in the Burrows-Wheeler Transform}
\titlerunning{On Undetected Redundancy in the Burrows-Wheeler Transform}
\author{Uwe Baier}{%
Institute of Theoretical Computer Science, Ulm University\\
{D-89069 Ulm, Germany}%
}{%
uwe.baier@uni-ulm.de%
}{%
https://orcid.org/0000-0002-0145-0332%
}{}
\authorrunning{U. Baier}
\Copyright{Uwe Baier}
\subjclass{ \\
\ccsdesc[500]{Theory of computation~Data compression} \\
\ccsdesc[300]{Applied computing~Document analysis} \\
\ccsdesc[100]{Mathematics of computing~Coding theory}
}
\keywords{Lossless data compression -- BWT -- Tunneling}
\supplement{Implementation available at \url{https://github.com/waYne1337/tbwt}.}

\EventEditors{Gonzalo Navarro, David Sankoff, and Binhai Zhu}
\EventNoEds{3}
\EventLongTitle{29th Annual Symposium on Combinatorial Pattern Matching (CPM 2018)}
\EventShortTitle{CPM 2018}
\EventAcronym{CPM}
\EventYear{2018}
\EventDate{July 2--4, 2018}
\EventLocation{Qingdao, China}
\EventLogo{}
\SeriesVolume{105}
\ArticleNo{3}
\nolinenumbers 
\hideLIPIcs  

\pgfplotsset{compat=1.9}
\usepgfplotslibrary{statistics}
\usetikzlibrary{backgrounds,calc,patterns,positioning,shadings}

\SetAlCapFnt{\bfseries \sffamily}
\renewcommand{\algorithmcfname}{\kern0.05em{\color[rgb]{0.99,0.78,0.07}\rule{0.73em}{0.73em}}%
\hspace*{0.67em}Algorithm}

\newcommand{\sdots}{..}
\newcommand{\SA}{\mathsf{SA}}
\newcommand{\BWT}{\mathsf{BWT}}
\newcommand{\llex}{<_{\mathsf{lex}}}
\newcommand{\rank}{\mathsf{rank}}
\newcommand{\select}{\mathsf{select}}
\newcommand{\C}{\mathsf{C}}
\newcommand{\LF}{\mathsf{LF}}
\newcommand{\LCol}{\mathsf{L}}
\newcommand{\FCol}{\mathsf{F}}

\newcommand{\outU}[1]{{#1}_{\mathsf{out}}}
\newcommand{\innU}[1]{{#1}_{\mathsf{in}}}
\newcommand{\cntF}{\mathsf{cntF}}
\newcommand{\cntL}{\mathsf{cntL}}
\newcommand{\AUX}{\mathsf{aux}}
\newcommand{\RB}{\mathcal R \mathcal B}

\newcommand{\conforms}{\hat{=}}
\newcommand{\indicator}[1]{\mathbf{1}_{#1}}
\newcommand{\rlencode}{\mathtt{rlencode}}

\begin{document}
\maketitle

\begin{abstract}
The Burrows-Wheeler-Transform ($\BWT$) is an invertible permutation of a text known to
be highly compressible but also useful for sequence analysis, what makes the BWT
highly attractive for lossless data compression. In this paper, we present a new technique
to reduce the size of a BWT using its combinatorial properties, while keeping it
invertible. The technique can be applied to any $\BWT$-based compressor, and, as
experiments show, is able to reduce the encoding size by $8-16 \%$ on average and
up to $33-57 \%$ in the best cases (depending on the $\BWT$-compressor used), making
$\BWT$-based compressors competitive or even superior to today's best lossless compressors.
\end{abstract}

\section{Introduction} \label{sec:introduction}
Lossless data compression plays an important role in modern digitization, as
it enables us to shift and save computation resources during information exchange.
For example, consider a setting where a computationally strong computer has to
distribute data over a limited channel--by use of data compression, the storage
requires less resources, and the file can be transmitted faster due to
reduced data size--with the drawback of extra computation time for en- and decoding
the information. Data compression is widespread today, current challenges
are not only to compress data, but also to serve special features like
resource-efficient decompression or even working on the compressed data directly,
because the only way to fit it in memory (and thus process it fast) consists of using
a compressed representation.

Compressors for the first mentioned feature typical make use of LZ77
\cite{ZIV:LEM:1977}--a compression technique that, briefly speaking, replaces
repeats in a text by references--resulting in very good compression rates and
very fast decompression. Popular examples of compressors using LZ77 are \texttt{gzip}
\cite{gzip-compressor} or \texttt{7-zip} \cite{7zip-compressor}, which can be
categorized as file transmission compressors. 

A different technique for the second mentioned feature makes use of
the Burrows-Wheeler-Transform ($\BWT$) \cite{BUR:WHE:1994}, which is an invertible
permutation of the characters in the original text. The $\BWT$ itself does not
compress data, but the transformed string tends to have some properties which make
it highly compressible. The most popular compressor of such kind is \texttt{bzip2}
\cite{bzip2-compressor}, but compression rates are not the only aspect that make the
$\BWT$ interesting: also, the $\BWT$ in combination with wavelet trees \cite{GRO:GUP:VIT:2003}
is well known to be an extremely useful and efficient tool for sequence analysis, commonly
known as FM-index \cite{FER:MAN:2005}.

As the $\BWT$ has very interesting combinatorial
properties, it has been heavily studied during the last two decades. For pure compression,
a summary of the most relevant ideas can be found in \cite{ABEL:2010}. Surprisingly, as Fenwick
\cite{FEN:2007} stated, less attention was given to the encoding of runs (a run 
is a continuous sequence of the same character). This was reasonable, because,
also stated by Fenwick \cite{FEN:2007}, \enquote{the original scheme proposed by Wheeler is
extremely efficient and unlikely to be much improved.}.

Finally, this is the point where our paper draws on. In this paper, we describe a new
technique called ``tunneling'', which relies on observations about combinatorics of a $\BWT$
and can be proved to maintain the invertibility (Theorem \ref{thm:tbwtinvertibility}).
The technique is independent from the way a run gets
encoded, but reduces the size of the $\BWT$ to be encoded eminently by shortening runs.
In numbers, we are able to reduce the encoding size of a compressed $\BWT$ by
$8-16 \%$ on average and up to $33-57 \%$ in the best cases, depending on the backend used
to compress a $\BWT$. This not only makes $\BWT$ compressors
competitive with today's best compressors, but also leaves the combinatorial
properties of the $\BWT$ intact, what indicates that the technique is
applicable to index structures such as the above mentioned FM-index%
\footnote{We give some hints towards this goal; there exist
	however more technical problems to be solved, outreaching the scope
	of this paper.}.

This paper is organized as follows: Section \ref{sec:preliminaries} contains basics
about the $\BWT$ and $\BWT$ compression. Section \ref{sec:tunneling} presents our new technique,
followed by a proof of invertibility of our representation in Section
\ref{sec:tunneling:invertibility}. Section \ref{sec:practicalimplementation} shows a way
to implement the technique, which is followed by experimental results in Section
\ref{sec:expresults} and conclusions in Section \ref{sec:conclusion}.

\section{Preliminaries} \label{sec:preliminaries}
First we want to describe the major parts of the $\BWT$ and its use in data compression.
Throughout this Paper, any interval $[i,j]$ or $[i,j)$ is meant to be an
interval over the natural numbers, every logarithm is of base 2, and indices start with $1$,
except when stated differently.

Let $\Sigma$ be a totally ordered set (alphabet) of elements (characters).
A string $S$ of length $n$ over alphabet $\Sigma$ is a finite sequence of $n$ 
characters originating from $\Sigma$. We call $S$ nullterminated if it ends with
the lowest ordered character $\mathtt{\$} \in \Sigma$ occurring only at the end of $S$.
The empty string with length $0$ is denoted by $\varepsilon$. Unless stated differently,
we assume that $S$ is nullterminated. Let $S$ be a string of length $n$, and let
$i,j \in [1,n]$. We denote by
\begin{itemize}
\item	$S[i]$ the $i$-th character of $S$.
\item	$S[i\sdots j]$ the substring of $S$ starting at the $i$-th and ending at the $j$-th position. \newline
	We state $S[i\sdots j] = \varepsilon$ if $i > j$, and define $S[i\sdots j) \coloneqq S[i\sdots j-1]$.
\item	$S_i$ the suffix of $S$ starting at the $i$-th position, i.e.\, $S_i = S[i\sdots n]$.
\item	$S_i \llex S_j$ if the suffix $S_i$ is lexicographically smaller than $S_j$,\newline
	i.e.\, there exists a $k \geq 0$ with $S[i\sdots i+k) = S[j\sdots j+k)$ and $S[i+k] < S[j+k]$.
\end{itemize}

\begin{definition} \label{def:suffarray}
Let $S$ be a string of length $n$.
The \textit{suffix array} \cite{MAN:MYE:1993} $\SA$ of $S$ is a permutation of integers in range 
$[1,n]$ satisfying ${S_{\SA[1]} \llex S_{\SA[2]} \llex \cdots \llex S_{\SA[n]}}$.
\end{definition}

\begin{definition} \label{def:burrowswheelertransform}
Let $S$ be a string of length $n$, and $\SA$ be its
corresponding suffix array. The Burrows-Wheeler-Transform ($\BWT$) of $S$
is a string $\LCol$ of length $n$ defined as $\LCol[i] \coloneqq S[\SA[i] - 1]$
if $\SA[i] > 1$ and $\LCol[i] \coloneqq \mathtt{\$}$ if $\SA[i] = 1$.
Also, we define the F-Column $\FCol$ as a string of length $n$ by $\FCol[i] \coloneqq S[\SA[i]]$,
which can also be obtained by sorting the characters in $\LCol$.
\end{definition}

In other words, the Burrows-Wheeler Transform is the concatenation of
characters which cyclically precede suffixes in the suffix array. An example
of a suffix array and a $\BWT$ can be found in Figure 
\ref{fig:samplesabwtrle}. An important property of the $\BWT$ is its invertibility, i.e.\, it's
possible to reconstruct the original string $S$ solely from its $\BWT$.
Therefore, we use the notation $\rank_{S}(c, i)$ to denote the number of
occurrences of character $c$ in the string $S[1\sdots i]$,
$\select_{S}(c, i)$ \nolinebreak to denote the position of the $i$-th occurrence of $c$ in $S$ and 
$\C_S[c]$ to denote the number of characters smaller
than $c$ in $S$, that is, $\C_S[c] \coloneqq \left| \{ ~ i \in [1,n] ~ | ~ S[i] < c ~ \} \right|$.

\begin{definition} \label{def:lfmapping}
Let $S$ be a string of length $n$, $\SA$ and $\LCol$ be its corresponding
suffix array and $\BWT$. The LF-mapping is a
permutation of integers in range $[1,n]$ defined as follows:
\[ \LF[i] \coloneqq \C_\LCol[\LCol[i]] + \rank_\LCol(\LCol[i], i) \]
We write $\LF^x[i]$ for the x-fold application of $\LF$, i.e.\,
${\LF^x[i] \coloneqq \underbrace{\LF[\LF[\dotsm\LF[}_{x \text{ times}}i]\dotsm]]}$,
\\[-\baselineskip] and define $\LF^0[i] \coloneqq i$.
\end{definition}

The LF-mapping carries its name because it maps each character in
$\LCol$ to its corresponding position in $\FCol$. To put it differently,
the LF-mapping induces a walk through the suffix array in reverse text
order, which commonly is called a ``backward step''.

\begin{lemma} \label{lem:bwtinvertibility}
Let $S$ be a string of length $n$, $\SA$ and $\LF$ be its
suffix array and LF-mapping. Then, $\SA[\LF[i]] = \SA[i] - 1$ holds
for all $i \in [1,n]$ with $\SA[i] \neq 1$. %
\hfill\begin{minipage}{.21\textwidth}
\begin{proof} See \cite{BUR:WHE:1994}.
\end{proof}
\end{minipage}
\end{lemma}

Thus, any $\BWT$ can be inverted by computing $\LF$ (which can solely be done using $\LCol$),
taking a walk through the suffix array in reverse text order using $\LF$ and
meanwhile collecting characters from $\LCol$, what yields the reverse string of $S$
(see \cite{BUR:WHE:1994} for more details). Our next concern
of the LF-mapping which will be important later is its parallelism property
inside runs, that is, a consecutive sequence of the same character in the $\BWT$.

\begin{lemma} \label{lem:lfrunparallelism}
Let $S$ be a string of length $n$, $\LCol$ and $\LF$ be its corresponding $\BWT$
and LF-mapping. For any interval $[i,j] \subseteq [1,n]$ with
$L[i] = L[i+1] = \dotsm = L[j]$, $\LF[i]+k = \LF[i+k]$ holds for all
$0 \leq k \leq j-i$. %
\hfill\begin{minipage}{.5\textwidth}
\begin{proof} Follows directly from Definition \ref{def:lfmapping}.
\end{proof}
\end{minipage}
\end{lemma}

\begin{figure}[t!]
\centering
\renewcommand{\arraystretch}{1.25}
\setlength{\tabcolsep}{0.5em}
\newcommand{\iatext}[2]{
	\textcolor{gray!50}{#1}\textcolor{black}{#2}
}
\scriptsize
\begin{tabular}{>{\scriptsize$}r<{$}|
                >{\scriptsize$}r<{$}||
                >{\scriptsize$\mathtt\bgroup}r<{\egroup$}|
                >{\scriptsize$\mathtt\bgroup}l<{\egroup$}||
                >{\scriptsize$\mathtt\bgroup}c<{\egroup$}|
                >{\scriptsize$\mathtt\bgroup}c<{\egroup$}||
                >{\scriptsize$\mathtt\bgroup}l<{\egroup$}}
i & \SA[i] 
  & \egroup S[1\sdots \SA[i]) \bgroup
  & \egroup S_{\SA[i]} \bgroup
  & \egroup \LCol[i] \bgroup 
  & \egroup \FCol[i] \bgroup
  & \egroup \rlencode(\LCol)[i] \bgroup \\ \hline

1	& 10	& \iatext{easypeas}{y}		& \$		& y  & \$ & y \\
2	& 7	& \iatext{easyp}{e}		& asy\$		& e  & a  & e \\
3	& 2	& \iatext{}{e}			& asypeasy\$	& e  & a  & 0 \\
4	& 6	& \iatext{easy}{p}		& easy\$	& p  & e  & p \\
5	& 1	& \egroup\varepsilon\bgroup	& easypeasy\$	& \$ & e  & \$ \\
6	& 5	& \iatext{eas}{y}		& peasy\$	& y  & p  & y \\
7	& 8	& \iatext{easyp}{ea}		& sy\$		& a  & s  & a \\
8	& 3	& \iatext{}{ea}			& sypeasy\$	& a  & s  & 0 \\
9	& 9	& \iatext{easyp}{eas}		& y\$		& s  & y  & s \\
10	& 4	& \iatext{}{eas}		& ypeasy\$	& s  & y  & 0 \\
\end{tabular}
\caption{Suffix array, Burrows-Wheeler-Transformation $\LCol$, $\FCol$, 
	run-length encoded $\BWT$ $\rlencode(\LCol)$
	and the prefixes preceding a
	suffix (third column) for ${S=\mathtt{easypeasy\$}}$. The $\BWT$
	is almost the same string as the concatenation of the last characters
	from prefixes preceding suffixes, except for the sentinel character.
	Also one can see that those prefixes often share more than one 
	character with the prefixes standing next to them.}
\label{fig:samplesabwtrle}
\end{figure}

To get an understanding why the $\BWT$ is useful for data compression,
we need a better understanding of it. The suffix array
 places lexicographically similar suffixes next to each other. Therefore, 
suffixes in subsequences of the suffix array often share a common prefix
(context). As the $\BWT$ consists of the cyclic preceding characters of
those suffixes, a subsequence of the $\BWT$ can be seen as a collection of 
characters preceding the same context in $S$. As a result, the character
distribution inside a subsequence of the $\BWT$ gets skew, i.e.\, it is dominated
by just a few characters which frequently appear before the context.

Typical $\BWT$ compressors make use of this fact by transforming the $\BWT$ such
that the locally skew character distributions turn into a global skew character
distribution--an example for such a transformation is given by the Move-To-Front
Transformation \cite{RYA:1980,BUR:WHE:1994}. Finally, the global skew character
distribution of the transform is useful for the last stage of typical $\BWT$
compressors: entropy encoding. The target of entropy encoding is
to minimize the middle cost for the encoding of a character in a string using its
character distribution. A well-known lower bound for this cost is given by
the entropy definition of Shannon \cite{SHA:1948}:

\begin{definition} \label{def:entropy}
Let $S$ be a string of length $n$, and let $\mathsf{c}_c$ be the count of 
character $c$ in $S$. The entropy $H(S)$ of string $S$ is defined as
$ H(S) ~\coloneqq~ \sum_{c \in \Sigma} \frac{\mathsf{c}_c}{n} \log \frac{n}{\mathsf{c}_c} 
   ~=~ \log n - \frac{1}{n} \sum_{c \in \Sigma} \mathsf{c}_c \log \mathsf{c}_c $.
\end{definition}

Over the years, a couple of methods were developed to
achieve cost-optimal entropy coding; most famous of such methods are Huffman
coding \cite{HUF:1952} and Arithmetic coding \cite{RIS:LAN:1979}. However,
 it can be shown that the more skew a character distribution of an underlying
source is, the more the entropy decreases. Consequently, the $\BWT$ transformation
normally is highly compressible using an entropy encoder.

Another trick for improving compression used by most state-of-the-art
$\BWT$ compressors is run-length-encoding. First of all, a run 
is a (length-maximal) subsequence in which all characters are equal. Run-length-encoding
transforms a run with height (length) $h$ of character $c$ into the string
$c\,h_k h_{k-1} \dotsm h_1$, where $(1\,h_k h_{k-1} \dotsm h_1)_2$ is the binary
representation of $h$ (in the encoding, the leading one is cut, and the symbols
$\mathtt{0}$ and $\mathtt{1}$ are assumed to be distinct from symbols in $S$). Figure
\ref{fig:samplesabwtrle} shows an example of run-length-encoding which often reduces
the length of a $\BWT$ drastically. We refer to 
\cite{FER:GIA:MAN:2006} for a survey about further $\BWT$ compression methods, and
introduce our last preliminary definition: the indicator function.

\begin{definition} \label{def:indicatorfunction}
Let $P$ be any boolean predicate. The indicator function $\indicator{P}$
then is defined as $\indicator{P} \coloneqq 1$ if predicate $P$ is true,
and $\indicator{P} \coloneqq 0$ otherwise.
\end{definition}

\section{Tunneling} \label{sec:tunneling}
The last section explained why the $\BWT$ produces long runs of the same
character. Briefly speaking, the consecutive characters in the $\BWT$ are
followed by similar contexts (suffixes) in the original text, and similar
contexts tend to be preceded by the same character. The $\BWT$ limits the
strings preceding contexts to just 1 character, but there is no reason why
longer preceding strings shouldn't be similar too%
\footnote{A similar observation was made in the context of self-repetitions in suffix arrays;
	see \cite{MAE:2000,NAV:MAE:2007}.
}. In fact, this
often is the case in repetitive texts: In Figure \ref{fig:samplesabwtrle},
the suffixes $S_{\SA[9]}$ and $S_{\SA[10]}$ are both preceded by the same
string $\mathtt{eas}$.

The intention of this section is to show a way how to use the similarity of the
preceding strings to reduce the size of the $\BWT$ while keeping the
invertibility and combinatorial properties of the $\BWT$ intact. Unlike existing
approaches \cite{MOF:ISA:2005}, we will not use word substitutions to achieve
this goal; the problem of word substitutions is to find a good substitution
scheme, as well as storing the word dictionary efficiently. Instead, we present
a method based on the combinatorics of the $\BWT$, which contains the
\enquote{word dictionary} implicitly in the remaining $\BWT$, and offers an easier
way to find a good substitution scheme--completely without substitution.
Our first step will be the definition of blocks, which are--short speaking--%
repetitions of the same preceding string in lexicographically consecutive suffixes.

\begin{definition} \label{def:blocks}
Let $S$ be a string of length $n$ and $\LCol$ be its $\BWT$. A block $B$ is a pair
of an integer $d$ and an interval $[i,j] \subseteq [1,n]$ ($d-[i,j]$-block) such that
\[ \LCol\left[\LF^x[i]\right] = \LCol\left[\LF^x[i+1]\right] = \dotsm =
   \LCol\left[\LF^x[j]\right] \text{ for all } 0 \leq x \leq d \]
We call $[i,j]$ the start interval of $B$, $\left[\LF^d[i],\LF^d[j]\right]$
the end interval of $B$, ${h_B \coloneqq j-i+1}$ the height of $B$ 
and $w_B \coloneqq d+1$ the width of $B$.
\end{definition}

\begin{figure}[h!]
\tikzset{x=1.25em,y=1.5ex,font=\scriptsize,
         every node/.style={minimum height=1.5ex,inner sep=1pt},
         intersection/.style={pattern=north east lines,pattern color=blue!75}
}
\centering
\begin{subfigure}[b]{.3\textwidth}
	\centering
	\begin{tikzpicture}
	\foreach \i/\j/\c [count=\k from 1] in {2/5/b,2/5/l,1/7/o,1/7/c,2/5/k} {
		\foreach \l in {\i,...,\j} {
			\node (n\l-\k) at (\k,-\l) {$\mathtt{\c}$};
		}
	}
	\draw (n1-3.north west) rectangle (n7-4.south east);
	\draw (n2-1.north west) rectangle (n5-5.south east);

	\begin{scope}[on background layer]
	\fill[intersection] (n2-3.north west) rectangle (n5-4.south east);
	\end{scope}
	\end{tikzpicture}
	\caption{compensable collision}
	\label{fig:blockcollisions:compensable}
\end{subfigure}
~
\begin{subfigure}[b]{.3\textwidth}
	\centering
	\begin{tikzpicture}
	\foreach \i/\j/\c [count=\k from 1] in {1/7/b,1/7/l,2/5/o,2/5/c,2/5/k} {
		\foreach \l in {\i,...,\j} {
			\node (n\l-\k) at (\k,-\l) {$\mathtt{\c}$};
		}
	}
	\draw (n1-1.north west) rectangle (n7-2.south east);
	\draw (n2-1.north west) rectangle (n5-5.south east);

	\begin{scope}[on background layer]
	\fill[intersection] (n2-1.north west) rectangle (n5-2.south east);
	\end{scope}
	\end{tikzpicture}
	\caption{critical collision}
	\label{fig:blockcollisions:critical}
\end{subfigure}
~
\begin{subfigure}[b]{.3\textwidth}
	\centering
	\begin{tikzpicture}
	\foreach \i in {1,2,...,7}
		\node[] (n\i) at (0,\i) {$\mathtt{a}$};

	\begin{scope}[on background layer]
	\filldraw[left color=gray!10,right color=gray!10,middle color=gray!20] 
	      ($(n7) - (1,.25)$)            to [out=90,in=150]
	      ($(n7.north) + (1,.35)$)
	      -- +(0,-5)                    to [out=120,in=90]
	      ($(n2) - (1,.25)$) -- cycle;
	\filldraw[left color=gray!5,right color=gray!5,middle color=white,line join=bevel] 
	      (n3.south)                    to [out=180,in=270]
	      ($(n4) - (1,.25)$) -- +(0,3)  to [out=270,in=180]
	      (n5.north)                    to [out=0,in=240]
	      ($(n5.north) + (1,.35)$) -- 
	      +(0,-3)                       to [out=240,in=0]
	      (n3.south);

	\filldraw[left color=gray!5,right color=white,line join=bevel] 
	      (n1.south east) -- (n1.south) to[out=180,in=270]
	      ($(n2) - (1,.25)$) -- +(0,3)  to[out=270,in=180]
	      (n3.north) -- (n3.north east) -- cycle;

	\filldraw[left color=white,right color=gray!5,line join=bevel] 
	      (n7.north)                    to [out=0,in=250]
	      ($(n7.north) + (1,.35)$) -- 
	      +(0,-3)                       to [out=250,in=0]
	      (n5.south) -- (n5.south west) -- (n7.north west) -- cycle;

	\fill[intersection] (n3.north east) rectangle (n3.south west);
	\fill[intersection] (n5.north east) rectangle (n5.south west);
	\end{scope}
	\end{tikzpicture}
	\caption{critical self-collision}
	\label{fig:blockcollisions:criticalself}
\end{subfigure}
\caption{Visualization of block collisions. Blocks are displayed as
	continuous stripes, shared positions are marked using diagonal lines.}
\label{fig:blockcollisions}
\end{figure}
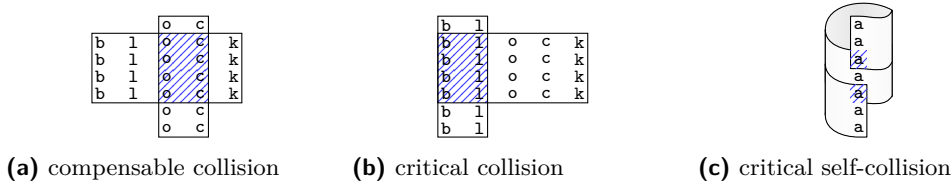

Blocks can also be seen as character matrices where each column consists of the
same character and each row is build by picking characters in $\BWT$ during $d$
consecutive applications of the LF-mapping, see also Figure \ref{fig:blockcollisions}.
An example of blocks can be found in Figure \ref{fig:samplesabwtrle}:
exemplary blocks are $2-[9,10]$, $1-[7,8]$, $0-[2,3]$ or $0-[5,5]$. We also
want to note that each column of a block can be mapped to a substring in the
$\BWT$ which consists of the same character.

\begin{definition} \label{def:tunnelingprocess}
Let $S$ be a string of length $n$, $\LCol$ and $\FCol$ be its corresponding
$\BWT$ and F-Column, and let $B = d-[i,j]$ be a
block of $S$. The process of tunneling block $B$ is defined as follows:
\begin{enumerate}
\item	cross out position $\LF^x[k]$ in $\LCol$ (mark it in a bitvector $\cntL$ of size $n$) \newline
	for all $0 \leq x < d$ and $i < k \leq j$.
\item	cross out position $\LF^x[k]$ in $\FCol$ (mark it in a bitvector $\cntF$ of size $n$) \newline
	for all $0 < x \leq d$ and $i < k \leq j$.
\item	remove positions $k$ that were crossed out both in $\LCol$ and $\FCol$ 
	from $\LCol$, $\FCol$ and the bitvectors $\cntL$ and $\cntF$.
\end{enumerate}
To tunnel a whole set $\mathcal B$ of blocks, we apply each step to
all blocks $B \in \mathcal B$ before continuing with the next step,
where the result ($\LCol$, $\cntL$ and $\cntF$) is called a tunneled $\BWT$.
\end{definition}

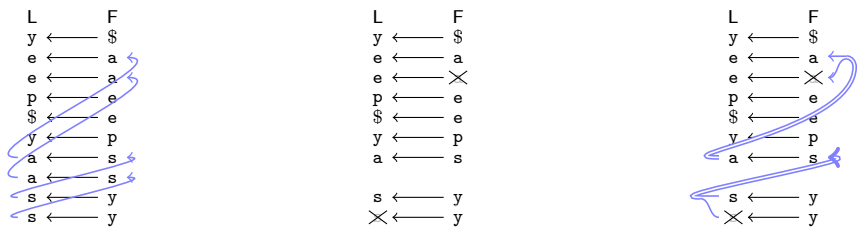
\begin{figure}[h!]
\tikzset{x=3em,y=1.75ex,node distance=1pt,font=\scriptsize,
	every node/.style={font=\scriptsize\vphantom{Ag\$},inner ysep=1pt,
	                   fill=white,fill opacity=0.5,text opacity=1},
	lfmapping/.style={semithick,->,blue!50},
	crossout/.style={fill=white,fill opacity=.75,text opacity=1}
}
\centering
\begin{subfigure}[t]{.3\textwidth}
	\centering
	\begin{tikzpicture}
	\node (l0) at (0,0) {$\LCol$};
	\node (f0) at (1,0) {$\FCol$};
	\foreach \l/\f [count=\i from 1] in {
			y/\$,e/a,e/a,p/e,\$/e,y/p,a/s,a/s,s/y,s/y} {
		\node (l\i) at (0,-\i) {$\mathtt{\l}$};
		\node (f\i) at (1,-\i) {$\mathtt{\f}$};
	};
	\foreach \lf in {1,...,10}
		\draw[->] (f\lf) -- (l\lf);
	\foreach \l/\f in {9/7,10/8,7/2,8/3}
		\draw[lfmapping] (l\l) to [out=180,in=0] (f\f);
	\node at ($(l1.west)-(10pt,0)$) {\phantom{a}};
	\node at ($(f1.east)+(10pt,0)$) {\phantom{a}};
	\end{tikzpicture}
	\caption{determine the block in the $\BWT$}
\end{subfigure}%
\quad%
\begin{subfigure}[t]{.3\textwidth}
	\centering
	\begin{tikzpicture}
	\node (l0) at (0,0) {$\LCol$};
	\node (f0) at (1,0) {$\FCol$};
	\foreach \l/\f [count=\i from 1] in {
			y/\$,e/a,e/a,p/e,\$/e,y/p,a/s,\phantom{a}/\phantom{s},s/y,s/y} {
		\node (l\i) at (0,-\i) {$\mathtt{\l}$};
		\node (f\i) at (1,-\i) {$\mathtt{\f}$};
	};
	\foreach \lf in {1,...,7,9,10}
		\draw[->] (f\lf) -- (l\lf);
	\node[crossout] at (l10) {\phantom{a}};
	\draw ($(l10.north west) + (2pt,-2pt)$) -- ($(l10.south east) - (2pt,-2pt)$);
	\draw ($(l10.south west) + (2pt,2pt)$) -- ($(l10.north east) - (2pt,2pt)$);
	\node[crossout] at (f3) {\phantom{a}};
	\draw ($(f3.north west) + (2pt,-2pt)$) -- ($(f3.south east) - (2pt,-2pt)$);
	\draw ($(f3.south west) + (2pt,2pt)$) -- ($(f3.north east) - (2pt,2pt)$);
	\node at ($(l1.west)-(10pt,0)$) {\phantom{a}};
	\node at ($(f1.east)+(10pt,0)$) {\phantom{a}};
	\end{tikzpicture}
	\caption{cross out positions and remove doubly crossed-out ones}
\end{subfigure}%
\quad%
\begin{subfigure}[t]{.3\textwidth}
	\centering
	\begin{tikzpicture}
	\node (l0) at (0,0) {$\LCol$};
	\node (f0) at (1,0) {$\FCol$};
	\foreach \l/\f [count=\i from 1] in {
			y/\$,e/a,e/a,p/e,\$/e,y/p,a/s,\phantom{a}/\phantom{s},s/y,s/y} {
		\node (l\i) at (0,-\i) {$\mathtt{\l}$};
		\node (f\i) at (1,-\i) {$\mathtt{\f}$};
	};
	\foreach \lf in {1,...,7,9,10}
		\draw[->] (f\lf) -- (l\lf);
	\node[crossout] at (l10) {\phantom{a}};
	\draw ($(l10.north west) + (2pt,-2pt)$) -- ($(l10.south east) - (2pt,-2pt)$);
	\draw ($(l10.south west) + (2pt,2pt)$) -- ($(l10.north east) - (2pt,2pt)$);
	\node[crossout] at (f3) {\phantom{a}};
	\draw ($(f3.north west) + (2pt,-2pt)$) -- ($(f3.south east) - (2pt,-2pt)$);
	\draw ($(f3.south west) + (2pt,2pt)$) -- ($(f3.north east) - (2pt,2pt)$);

	\draw[lfmapping,-] (l10.west) to [out=180,in=0] ($(l9.west)-(7pt,.5pt)$);
	\draw[lfmapping,-] ($(l9.west)+(0,.5pt)$) -- ($(l9.west)-(7pt,-.5pt)$);
	\draw[lfmapping,double,arrows= {->}] ($(l9.west)-(7pt,0)$) to [out=180,in=0] (f7.east);
	\draw[lfmapping,double,-] (l7.west) to [out=180,in=270] ($(f3.east)+(10pt,4pt)$) to [out=90,in=0] ($(f2.east)+(7pt,0)$);
	\draw[lfmapping] ($(f2.east)+(7pt,.5pt)$) -- ($(f2.east)+(0,.5pt)$);
	\draw[lfmapping] ($(f2.east)+(7pt,-.5pt)$) to [out=180,in=0] (f3.east);

	\node at ($(l1.west)-(10pt,0)$) {\phantom{a}};
	\node at ($(f1.east)+(10pt,0)$) {\phantom{a}};
	\end{tikzpicture}
	\caption{Reconstruction idea: use one block row for both rows}
	\label{fig:tunnelingprocess:reconstructidea}
\end{subfigure}
\caption{Process of tunneling as described in Definition \ref{def:tunnelingprocess}.
	Above, block $2-[9,10]$ from the running example is tunneled. Any lines colored
	blue are related to the LF-mapping, cross-outs in $\LCol$ and $\FCol$ are displayed
	by crosses.
}
\label{fig:tunnelingprocess}
\end{figure}

A simpler description of tunneling a block can be given as follows:
Remove all positions of the block from the $\BWT$, except for those
in the rightmost or leftmost column or uppermost row of the block. Afterwards,
cross out positions in the interval start from $\LCol$ and the interval end from $\FCol$,
both except for the uppermost position. An example can be seen in Figure
\ref{fig:tunnelingprocess}.

The interesting question will be if we are able to reconstruct the text, even
if we remove positions from the $\BWT$. Beforehand however, we need to care about
block intersections, since they can produce side effects when tunneling more than one
block.

\begin{definition} \label{def:blockcollisions}
Let $B = d-[i,j]$ and $\tilde B = \tilde d-[\tilde i,\tilde j]$ be two different blocks
of a string $S$ with $\BWT$ $\LCol$ and LF-mapping $\LF$. We say that $B$
and $\tilde B$ collide if they share positions in $\LCol$, i.e.\,
there exists $x \in [0,d\,]$ and $\tilde x \in [0,\tilde d\,]$ such that
\[
	\left[ \LF^{\vphantom{\tilde x}x}[i], \LF^{\vphantom{\tilde x}x}[j] \right] \cap
	\left[ \LF^{\tilde x}[\tilde i], \LF^{\tilde x}[\tilde j] \right]
	\neq \emptyset
\] 
We call each position in the intersection a shared position.
Analogously, we say that a block $B = d-[i,j]$ is self-colliding if
some $x,\tilde x \in [0,d]$ with $x < \tilde x$ exist such that
\[
	\left[ \LF^{\vphantom{\tilde x}x}[i], \LF^{\vphantom{\tilde x}x}[j] \right] \cap
	\left[ \LF^{\tilde x}[i], \LF^{\tilde x}[j] \right]
	\neq \emptyset
\]

Furthermore, let $\innU{B}$ and $\outU{B}$ be two colliding blocks and w.l.o.g.\
${h_{\innU{B}} \geq h_{\outU{B}}}$. We call the collision between $\innU{B}$ and $\outU{B}$
compensable if following conditions are fulfilled:
\begin{enumerate}
\item	\label{def:blockcollisions:compensable:firstlastunshared}
	The leftmost and rightmost columns of $\outU{B}$ do not intersect: \\
	The positions $\left[\outU{i},\outU{j}\right]$ and $\left[\LF^{\outU{d}}[\outU{i}],\LF^{\outU{d}}[\outU{j}]\right]$ are not shared

\item	\label{def:blockcollisions:compensable:rowunshared}
	At least one row of $\innU{B}$ does not intersect: \\
	There exists a $y \in [\innU{i},\innU{j}]$ such that the positions $y,\LF[y],\ldots,\LF^{\innU{d}}[y]$ are not shared

\item	\label{def:blockcollisions:compensable:probersubrow}
	The intersection area forms a block of height $\innU{B}-\outU{B}$: \\ 
	For all $\innU{x} \in [0,\innU{d}]$ and $\outU{x} \in [0,\outU{d}]$ following holds:
	\[ \bigg| ~
		\Big[ \LF^{\innU{x}}[\innU{i}], \LF^{\innU{x}}[\innU{j}] \Big] \setminus
		\Big[ \LF^{\outU{x}}[\outU{i}], \LF^{\outU{x}}[\outU{j}] \Big] ~
	   \bigg| \in \{ 0, h_{\innU{B}}-h_{\outU{B}} \} \]
\end{enumerate}
If the conditions are not fulfilled, or if a collision is a self-collision, we call it critical.
\end{definition}

Figure \ref{fig:blockcollisions} visualizes the different kinds of block collisions.
Examples of block collisions can be found in Figure \ref{fig:samplesabwtrle}:
the blocks $2-[9,9]$ and $0-[7,8]$ form a compensable collision, while
the blocks $2-[9,9]$ and $1-[7,8]$ form a critical collision.
Furthermore, for $\LCol = \mathtt{aa}\dotsm\mathtt{a\$}$ any block of width
and height greater than one is self-colliding.

Now let us discuss the effect of the classifying
criteria of collisions from Definition \ref{def:blockcollisions} 
a bit further. By the criteria, a compensable collision always consists
 of a ``wider'' (outer) and a ``shorter but higher'' (inner) block: The start and end interval
of the outer block contain no shared position (condition
\ref{def:blockcollisions:compensable:firstlastunshared}), which in conjunction with
condition \ref{def:blockcollisions:compensable:probersubrow} implies that the
outer block must be wider than the inner one. For the inner block, at least one row
must be unshared (condition \ref{def:blockcollisions:compensable:rowunshared}), what
in conjunction with condition \ref{def:blockcollisions:compensable:probersubrow} analogously
shows that the inner block must be higher than the outer one. In the sense of visualization,
these conditions build kind of a cross overlay of blocks, as depicted in
Figure \ref{fig:blockcollisions:compensable}.
Extending this idea to more than two blocks, the criteria forms a natural hierarchy
on colliding blocks, which will be useful for invertibility issues.

\section{Invertibility} \label{sec:tunneling:invertibility}
The purpose of this section is to show that a tunneled $\BWT$ can be inverted,
i.e.\, the original string from which a $\BWT$ was constructed from can be rebuilt.
As a first step, we will introduce a new generalized LF-mapping.

\begin{definition} \label{def:generalizedlf}
Let $\LCol$ be the Burrows-Wheeler-Transformation of a string of length $n$ and 
let $\cntL$ and $\cntF$ be two bitvectors of size $n$. The generalized
LF-mapping is defined by
\[ \LF_{\cntF}^{\cntL}[i]
	~\coloneqq~
	\underbrace{\vphantom{\sum_{k=1}^{i}} \select_{\cntF}(1}%
		_{^\text{skip removed}_\text{\quad positions}}
	~,~
	\underbrace{\vphantom{\sum_{k=1}^{i}} \sum_{k=1}^n \cntL[k] \cdot \indicator{\LCol[k] < \LCol[i]}}%
		_{\conforms ~ \C_{\LCol}[\LCol[i]]}
	~+~ 
	\underbrace{\vphantom{\sum_{k=1}^{i}} \sum_{k=1}^{i} \cntL[k] \cdot \indicator{\LCol[k] = \LCol[i]} }%
		_{\conforms ~ \rank_{\LCol}( \LCol[i], i)}
	~)
\]
\end{definition}

Definition \ref{def:generalizedlf} is almost equal to the normal LF-mapping--%
except that characters crossed out in $\LCol$ are ignored, while characters
crossed out in $\FCol$ are skipped. Next, we will see that this definition is reasonable
for tunneling, as it maintains its structure when removing positions from
$\LCol$ in the ``right'' manner.

\begin{lemma} \label{lem:rankpreserve}
Let $\LCol$ be the Burrows-Wheeler-Transformation of a string of length $n$ with LF-mapping $\LF$ and 
let $\cntL$ and $\cntF$ be two bitvectors of size $n$. Then following properties hold
for the generalized LF-mapping $\LF_{\cntF}^{\cntL}$:
\begin{enumerate}
\setlength\itemsep{1ex}
\item	Let $\cntF[i] = \cntL[i] = 1$ for all $i \in [1,n]$.
	Then, $\LF_{\cntF}^{\cntL}$ is identical to the normal LF-mapping $\LF$.
	\label{lemprop:rankpreserve:identicaltolf}
\item	Let $\cntF$ and $\cntL$ be two bitvectors such that $\LF_{\cntF}^{\cntL}[j] = \LF[j]$ for all
	$j$ with $\cntL[j] = 1$. Let $i$ be an integer with $\cntL[i] = \cntF\left[\LF[i]\right] = 1$.
	Crossing out position $i$ in $\LCol$ and position $\LF[i]$ in $\FCol$
	(setting $\cntL[i] = \cntF\left[\LF[i]\right] = 0$) does not change the mapping: Define
	\[
	\widetilde{\cntL}[j] \coloneqq \cntL[j] \cdot \indicator{j \neq i}
	\text{\quad and \quad}
	\widetilde{\cntF}[j] \coloneqq \cntF[j] \cdot \indicator{j \neq \LF[i]}
	\]
	Then, $\LF_{\widetilde{\cntF}}^{\widetilde{\cntL}}[j] = \LF[j]$ 
	for all $j$ with $\widetilde{\cntL}[j] = 1$.
	\label{lemprop:rankpreserve:clearingvalues}
\item	Let $i$ be an integer with $\cntL[i] = \cntF[i] = 0$. Removing position $i$
	(crossed out both in $\LCol$ and $\FCol$) from $\LCol$, $\cntL$ and $\cntF$ does
	not change the mapping: Define
	\[
	\widetilde{\cntL}[j] \coloneqq \cntL[j + \indicator{j \geq i}]
	\text{~,~}
	\widetilde{\cntF}[j] \coloneqq \cntF[j + \indicator{j \geq i}]
	\text{~and~}
	\widetilde{\LCol}[j] \coloneqq \LCol[j + \indicator{j \geq i}]
	\]
	Then, for the corresponding mapping $\widetilde{\LF}^{\widetilde{\cntL}}_{\widetilde{\cntF}}$ the following holds:
	\[
	\widetilde{\LF}^{\widetilde{\cntL}}_{\widetilde{\cntF}}[j] =
	\LF_{\cntF}^{\cntL}[j + \indicator{j \geq i}] - \indicator{\LF_{\cntF}^{\cntL}[j + \indicator{j \geq i}] \geq i}
	\]
	\label{lemprop:rankpreserve:removingcleared}
\end{enumerate}
\end{lemma}

Lemma \ref{lem:rankpreserve} looks really bulky at the first moment, but reflects the
operations required for tunneling, as we will see by discussing its different properties.
Property \ref{lemprop:rankpreserve:identicaltolf} says that the new LF-mapping is identical
to the old if we cross out nothing, and is straightforward. The more interesting property
\ref{lemprop:rankpreserve:clearingvalues} tells us that the LF-mapping stays
identical if we cross out consecutive positions in $\LCol$ and $\FCol$ in terms of text
order. To explain why this works, think of a character $c$ at position $i$ in a $\BWT$.
If we cross out $c$ in $\LCol$, it is ignored, and thus the LF-mapping of all characters
in $\LCol$ which are greater than $c$ (or equal to $c$ but placed below of $i$) gets
shifted one position upwards, and thus is modified. Now, as we also cross out $\LF[i]$ in
$\FCol$, and crossed-out positions in $\FCol$ are skipped, all of the modified positions are
shifted one position downward, because their original LF-mapping was greater than
$\LF[i]$, and thus, the mapping stays identical. Property
\ref{lemprop:rankpreserve:removingcleared} also is easy to understand, as it
 says that a position which is ignored in $\LCol$ and skipped in $\FCol$ can be
completely removed. A formal proof of the properties can be found in Appendix
\ref{sec:proofrankpreserve}, but for now let us concentrate on the reconstruction
of the original text.

The idea for reconstruction will be as follows: According to Lemma \ref{lem:rankpreserve},
tunneling will leave the LF-mapping identical, so we need only to clarify
how to deal with tunneled blocks. As a remainder, tunneling means that we remove
all of the characters except for the rightmost and leftmost column and uppermost row of a block.
In blocks, we know that each row of the block is identical, and all of the rows run
in parallel (in terms of the LF-mapping). Thus, if we reach the start of a tunnel,
we will save the offset to the uppermost row, proceed at the uppermost row, and, once
we leave the tunnel, use the saved offset to get back to the correct ``lane'', as shown
Figure \ref{fig:tunnelingprocess:reconstructidea} and described in Algorithm
\ref{alg:textreconstruction}.

\begin{algorithm}[h!]
\scriptsize
\KwData{Tunneled Burrows-Wheeler-Transform $\LCol$ of size $n$ from string $S$
	of size $\widetilde{n}$, bitvectors $\cntL$ and $\cntF$ of size $n+1$ with
	$\cntL[n+1] = \cntF[n+1] = 1$.}
\KwResult{String $S$ from which $\LCol$, $\cntL$ and $\cntF$ were build from.}
\SetKwFunction{push}{$s$.push}\SetKwFunction{pop}{$s$.pop}\SetKwFunction{top}{$s$.top}
\BlankLine
\DontPrintSemicolon
	let $\LF^*$ be the generalized LF-mapping (Definition \ref{def:generalizedlf}).
	\tcp*[r]{see Appendix \ref{sec:generalizedlfmappingcomputation} for computation}
	initialize an empty stack $s$\;
	$S\left[ \widetilde{n} \right] \gets \mathtt{\$}$\;
	$j \gets 1$\;
	\For{$i \gets \widetilde{n}-1$ \KwTo $1$}{
		\If(\tcp*[f]{end of a tunnel}){$\cntF[j+1] = 0$}{
			$j \gets j + \top{}$\;
			$\pop{}$\;
		}
		$S[i] \gets \LCol[j]$\; \label{algref:textreconstruction:charpickup}
		\If(\tcp*[f]{start of a tunnel}){$\cntL[j] = 0$ or $\cntL[j+1] = 0$}{
			$k \gets \max\{ l \in [1,j] \mid \cntL[l] = 1\}$ \label{algref:textreconstruction:getuppermostrow}
			\tcp*[r]{uppermost row of block}
			$\push{j - k}$\;
		}
		$j \gets \LF^*[j]$\; \label{algref:textreconstruction:lfwalk}
	}
\caption{Inverting a tunneled Burrows-Wheeler-Transform.}
\label{alg:textreconstruction}
\end{algorithm}

The reconstruction process also inspired us to name the method tunneling:
once the start of a block is reached, the offset to the uppermost row gets saved,
and we enter the ``tunnel'', namely the uppermost row. After the tunnel ended,
the temporarily information is used to get back to the correct ``lane'', that is
the row on which we entered the tunnel.

\begin{theorem} \label{thm:tbwtinvertibility}
Let $S$ be a string of length $\widetilde{n}$, let $\mathcal B$ be a
set of blocks in its $\BWT$ containing no critical block collisions,
and let $\LCol$, $\cntL$ and $\cntF$ (each of size $n$)
 be the components emerging by tunneling the blocks of set $\mathcal B$.
Then, Algorithm \ref{alg:textreconstruction} reconstructs the string $S$
from the tunneled $\BWT$ in $O(\widetilde{n})$ time.

\begin{proof}
First, note that the generalized LF-mapping in a tunneled $\BWT$ conforms
to the normal LF-mapping in a traditional $\BWT$: Definition
\ref{def:tunnelingprocess} tells us that for each marked position
$i$ in $\cntL$, the associated
position $\LF[i]$ is marked in $\cntF$ during tunneling process. Furthermore, tunneling
only removes positions $i$ with $\cntL[i] = \cntF[i] = 0$---note that this 
argumentation still is true for any colliding blocks, with only the only difference
that some of those ``position pairs'' are marked more than once. Thus Lemma \ref{lem:rankpreserve}
ensures that a walk through the generalized LF-mapping (Line
\ref{algref:textreconstruction:lfwalk} of Algorithm \ref{alg:textreconstruction}) 
will reproduce the same string during character pickup (Line
\ref{algref:textreconstruction:charpickup})---except for positions $i$ with either
$\cntL[i]$ or $\cntF[i]$ equal zero, which are positions in a start or end interval
of a block.

As we know that
each row of a block is identical (in terms of characters), and that the LF-mapping
in a block runs in parallel (Lemma \ref{lem:lfrunparallelism}), for correct reconstruction, it's
sufficient to store the relative offset to the top of a block when entering it,
reconstruct any of the rows of the block, and at the end of the block use the stored
offset to step back to the relative position on which the block was entered---as
performed by Algorithm \ref{alg:textreconstruction}. In case of block collisions,
by the hierarchy build from the condition of compensable collisions, a tunnel will
not be left until all its inner colliding block tunnels are left, thus the stack
in Algorithm \ref{alg:textreconstruction} correctly matches each tunnel end with
the offset the tunnel was entered.

Finally, Algorithm \ref{alg:textreconstruction} can be implemented to require
$O( \widetilde{n} )$ time by precomputing the generalized LF-mapping in 
$O(n)$ time (see Appendix \ref{sec:generalizedlfmappingcomputation}), and by
implementing line \ref{algref:textreconstruction:getuppermostrow} with an array
mapping each position to the nearest previous position $i$ with $\cntL[i] = 1$,
which obviously also can be computed in $O( n )$ time, requiring
$O(1)$ time per query.
\end{proof}
\end{theorem}

In contrast, when dealing with critical collisions, Algorithm \ref{alg:textreconstruction}
will not be able to correctly match a tunnel start or end to the corresponding tunnel
due to the intersections of start intervals or end intervals of blocks.
In the case of self-collisions, the tunneling process from Definition \ref{def:tunnelingprocess}
will remove entries of the topmost row of the self-colliding block from the $\BWT$, thus falsifying
the correctness proof of Algorithm \ref{alg:textreconstruction}. 

\section{Practical Implementation} \label{sec:practicalimplementation}
This section's purpose is to give a brief summary on how to use $\BWT$ tunneling
practically. Our first restriction therefore is to focus only on such-called
run-blocks, what will make tunneling easier to handle. A run-block is a block
whose start and end intervals have the same height as the runs in the $\BWT$
where the intervals occur. Furthermore, we will focus only on width-maximal
run-blocks having height and width both greater than one.

\begin{definition} \label{def:runblocks}
Let $S$ be a string of length $n$ and $\LCol$ be its $\BWT$. Furthermore, assume
that the border cases $\LCol[0]$ and $\LCol[n+1]$ contain characters such that
$\LCol[0] \neq \LCol[1]$ and $\LCol[n] \neq \LCol[n+1]$.

A $d-[i,j]$-block is called a run-block if
${\LCol[i] \neq \LCol[i-1]}$,
${\LCol[j] \neq \LCol[j+1]}$, \newline
${\LCol[\LF^d[i]] \neq \LCol[\LF^d[i] - 1]}$ and
${\LCol[\LF^d[j]] \neq \LCol[\LF^d[j] + 1]}$
 holds.

A run-block $RB$ is called width-maximal if it is wider than any colliding
run-block $\widetilde{RB}$ with same height, i.e.\ $RB$ and $\widetilde{RB}$
collide and $h_{B} = h_{\widetilde{RB}} \Rightarrow w_{RB} \geq w_{\widetilde{RB}}$.
\end{definition}

 An example
of run-blocks is given in Figure \ref{fig:samplesabwtrle}: $0-[2,3]$, $1-[7,8]$
are run-blocks, $2-[9,10]$ is the only width-maximal run-block. As a counter
example, $2-[9,9]$ is no run-block, as $\LCol[9] = \LCol[10]$, thus the height
is not identical to that of the run where the block starts.

Run-blocks with height greater than one will never be self-colliding%
\footnote{
	Self-collisions are related to overlapping repeats in the text.
	Thus, a self-colliding block has to share positions in its start interval with itself,
	what cannot happen in run-blocks: as the overlapping intervals 
	of a block cannot be equal ($\LF$ is a permutation), the start interval
	wouldn't be height-maximal.}%
--also, any collision between width-maximal run-blocks always is compensable: a run-block
is height-maximal in sense of its start- and end-interval, thus any collision enforces
one block to be higher and on the ``inside'' of the other, as required by Definition
\ref{def:blockcollisions}.

\subsection{Block Computation} \label{sec:practicalimplementation:blockcomputation}
Our first concern is how blocks can be computed. For arbitrary blocks, a simple
solution would be to compute the pairwise longest common suffixes of
$S[1\sdots \SA[i])$ and $S[1\sdots SA[i+1])$, and afterwards enumerate the blocks
using a stack-based approach--which is possible in $O(n)$, see
\cite{KAE:KEM:PUG:2012} and \cite{KAS:LEE:ARI:ARI:PAR:2001}. However, as we want
to compute the restricted set of width-maximal run-blocks, we will choose a different
approach.

We will describe the idea of the approach only; see Appendix \ref{sec:runblockcomputation}
for an algorithm. The idea is to use runs as a start point, and use the LF-mapping to proceed
over the $\BWT$. Every time a run is reached which allows to width-extend
the current block, the current block is pushed on a stack and the run
is used as new block. Then, as soon as the current block cannot be extended (because
the current run is not high enough), blocks are popped from the stack until an extendable block
is reached. During the removal of blocks, the necessary conditions for
width-maximal run-blocks can be checked. Also, to increase efficiency, a side-effect
similar to pointer jumping is used, which allows to skip already processed blocks.

\subsection{Tunneled $\BWT$ Encoding} \label{sec:practicalimplementation:tbwtencoding}
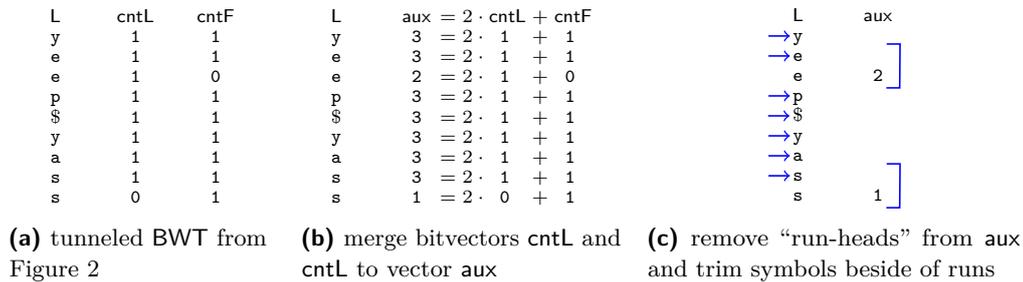
\begin{figure}[h!]
\tikzset{x=3em,y=1.75ex,font=\scriptsize,
         every node/.style={font=\scriptsize\vphantom{Ag\$},minimum height=1.5ex,inner sep=1pt},
         runsep/.style={dotted,semithick}
}
\centering
\begin{subfigure}[b]{.25\textwidth}
	\centering
	\begin{tikzpicture}
	\node (l0) at (0,0) {$\LCol$};
	\node (cl0) at (1,0) {$\cntL$};
	\node (cf0) at (2,0) {$\cntF$};

	\foreach \l [count=\i from 1] in {y,e,e,p,\$,y,a,s,s}
		\node (l\i) at (0,-\i) {$\mathtt{\l}$};

	\foreach \cl [count=\i from 1] in {1,1,1,1,1,1,1,1,0}
		\node (cl\i) at (1,-\i) {$\mathtt{\cl}$};

	\foreach \cf [count=\i from 1] in {1,1,0,1,1,1,1,1,1}
		\node (cf\i) at (2,-\i) {$\mathtt{\cf}$};
	\end{tikzpicture}
	\caption{tunneled $\BWT$ from \newline Figure \ref{fig:tunnelingprocess}}
\end{subfigure}
~
\begin{subfigure}[b]{.3\textwidth}
	\centering
	\begin{tikzpicture}
	\node (l0) at (0,0) {$\LCol$};
	\node (a0) at (1,0) {$\AUX$};
	\node[anchor=west] (cntL0) at (1.25,0) {$=2 \cdot \cntL$};
	\node[anchor=west] (cntF0) at (2.4,0) {$+\;\cntF$};

	\foreach \l [count=\i from 1] in {y,e,e,p,\$,y,a,s,s}
		\node (l\i) at (0,-\i) {$\mathtt{\l}$};

	\foreach \a [count=\i from 1] in {3,3,2,3,3,3,3,3,1}
		\node (a\i) at (1,-\i) {$\mathtt{\a}$};

	\foreach \cntl [count=\i from 1] in {1,1,1,1,1,1,1,1,0}
		\node[anchor=west] (cntL\i) at (1.25,-\i) {$=2 \cdot ~\,\mathtt{\cntl}$};
	\foreach \cntf [count=\i from 1] in {1,1,0,1,1,1,1,1,1}
		\node[anchor=west] (cntF\i) at (2.4,-\i) {$+\;~\,\mathtt{\cntf}$};
	\end{tikzpicture}
	\caption{merge bitvectors $\cntL$ and $\cntL$ to vector $\AUX$}
\end{subfigure}
~
\begin{subfigure}[b]{.35\textwidth}
	\centering
	\begin{tikzpicture}
	\node (l0) at (0,0) {$\LCol$};
	\node (a0) at (1,0) {$\AUX$};

	\foreach \l [count=\i from 1] in {y,e,e,p,\$,y,a,s,s}
		\node (l\i) at (0,-\i) {$\mathtt{\l}$};

	\foreach \a [count=\i from 1] in {\phantom{0},\phantom{0},2,\phantom{0},\phantom{0},
		                          \phantom{0},\phantom{0},\phantom{0},1}
		\node (a\i) at (1,-\i) {$\mathtt{\a}$};

	\foreach \rh in {1,2,4,5,6,7,8}
		\draw[blue,semithick,->] ($(l\rh.west) - (8pt,0)$) -- (l\rh.west);

	\foreach \rs/\re in {2/3,8/9}
		\draw[blue,semithick] (a\rs.north east) -- ($(a\rs.north east) + (5pt,0)$) --($(a\re.south east) + (5pt,0)$)  -- (a\re.south east);
	\end{tikzpicture}
	\caption{remove ``run-heads'' from $\AUX$ and trim symbols beside of runs}
\end{subfigure}
\caption{Exemplary $\AUX$-vector generation from a tunneled $\BWT$. ``Run-heads'' (first symbols of runs) are marked using arrows,
	runs with height greater 1 by right square brackets.}
\label{fig:runblocks-auxencoding}
\end{figure}

The encoding of a tunneled $\BWT$ requires to encode three components: the remaining $\BWT$ $\LCol$,
as well as the bitvectors $\cntL$ and $\cntF$. To reduce a component, we merge $\cntL$ and $\cntF$
to a new vector named $\AUX$ by setting $\AUX[i] \coloneqq 2 \cdot \cntL[i] + \cntF[i]$.
The vector $\AUX$ now contains 3 distinct symbols%
\footnote{Positions $i$ containing markings in both $\cntL[i]$ and $\cntF[i]$ were
	removed by tunneling.}%
, and is further shortened by removing all positions where runs in $\LCol$ start
(all of this positions must be unmarked as the topmost row of a run-block stays unchanged)
and by trimming the (identical) remaining symbols beside of each run of height greater
than one to just one symbol (reconstruction is possible as only run-blocks are tunneled).
An example is listed in Figure \ref{fig:runblocks-auxencoding}.

As the components $\LCol$ and $\AUX$ originate from the same source and are quite similar,
we'll encode both components with the same $\BWT$ backend encoder, like, for example,
Move-To-Front-Transformation + run-length-encoding of zeros + entropy encoding. This not
only simplifies the implementation, but also allows to uncouple the block choice from
the used backend encoder: As tunneling leaves the uppermost row of a block intact, the
effect of tunneling can be summarized as shortening runs in $\LCol$ at cost of increasing
the number of runs in $\AUX$ due to the tunnel-marking symbols. For a good block-choice,
it is useful to think of a tax system: a good choice maximizes the net-benefit,
which is given by gross-benefit (amount of information removed from $\LCol$) minus the
tax (amount of information required to encode $\AUX$). Now, as long as different backend
encoders encode runs in a similar fashion, an optimal block choice for a specific backend
encoder will be near-optimal for all the other backend encoders, as the efficiency of such
encoders can be seen as a constant $c$ which does not affect the maximization of the net-benefit.

As most state-of-the-art compressors use run-length-encoding, we estimate net-benefit
and tax in terms of run-length-encoding. Consequently, the net-benefit between a normal $\BWT$ $\LCol$
and a tunneled $\BWT$ $\widetilde{\LCol}$ is given by
\begin{equation}
\begin{aligned}
\FuncSty{gross-benefit} & \coloneqq
	|\rlencode(\LCol)|\cdot H(\rlencode(\LCol))
	- |\rlencode(\widetilde{\LCol})|\cdot H(\rlencode(\widetilde{\LCol})) \\
 & \approx
	n \log\left( \frac{n}{n-tc} \right)
	- rc \log\left( \frac{rc}{rc-tc} \right)
	+ tc \left( 1 + \log\left( \frac{n-tc}{rc-tc} \right)\right)
\end{aligned}
\label{eq:tunneling:gross-benefit}
\end{equation}
where $n \coloneqq |\rlencode(L)|$ is the number of characters of the run-length-encoding of $\LCol$,
$H$ is the entropy function from Definition \ref{def:entropy},
$rc$ is the number of run-characters in $\rlencode(\LCol)$ (all characters except for the run-heads) and
$tc$ is the number of removed run-characters in $\rlencode(\widetilde{\LCol})$
, see also Appendix \ref{sec:benefittaxestimators}.
The tax is given by
\begin{equation}
\begin{aligned} 
\FuncSty{tax} & \coloneqq
	|\rlencode(\AUX)| \cdot H(\rlencode(\AUX)) \\
 & \approx
	2\cdot t \cdot (1 + \log(h^2 - 1)) 
	+ 2\cdot t\cdot h \cdot \log\left(1 + \frac{2}{h-1} \right)
\end{aligned}
\label{eq:tunneling:tax}
\end{equation}
where $t$ is the number of tunneled Blocks and
$h \coloneqq \log\left( \frac{r_{h > 1} - 2\cdot t}{2\cdot t} \right)$
with $r_{h>1}$ being the number of runs with height greater than $1$, see also Appendix \ref{sec:benefittaxestimators}.


\subsection{Block Choice} \label{sec:practicalimplementation:blockchoice}
The estimators from the last section give a clear indication that tunneling will not
always improve compression--picking too small blocks may result in a tax whose size
overcomes the gross-benefit. It is clear however that bigger blocks are preferable
to smaller ones--thus a greedy approach will produce best results. Algorithm
\ref{alg:blockchoice} sketches our strategy for choosing blocks, also considering that
the gross-benefit and the tax do not grow in a likewise manner.

\newcommand{\nbbest}{{b}_{\mathsf{best}}}
\newcommand{\tbest}{t_{\mathsf{best}}}
\newcommand{\BSorted}{\mathsf{BS}}
\begin{algorithm}[ht]
\scriptsize
\SetKwFunction{score}{score}
\KwData{a set $\RB$ of width-maximal run-blocks and a function $\score$
	which for each block returns the amount of run-characters it removes.}
\KwResult{the array $\BSorted$ and a number $\tbest$, whereby $\BSorted[1..\tbest]$ contains the blocks of a greedy block choice.}
\BlankLine
\BlankLine
\DontPrintSemicolon
	let $\BSorted$ be an array of size $|\RB|$\;
	$\tbest \gets 0$ \tcp*[r]{number of tunnels allowing best benefit}
	$\nbbest \gets 0$ \tcp*[r]{best tunneling net-benefit in bits}
	$tc \gets 0$ \tcp*[r]{number of run-characters removed by tunneling}
	\For{$t\gets 1$ \KwTo $|\BSorted|$} {
		let $B \in \RB$ be the block with $\score{$B$}$ maximal\; \label{alg:blockchoice:maxScore}
		reduce $\score{$\tilde B$}$ for all colliding blocks $\tilde B$ of $B$ depending on the kind of collision\; \label{alg:blockchoice:reducescore}
		remove collisions between all inner and outer colliding blocks of $B$
		\tcp*[f]{intersecting area gets tunneled}\;  \label{alg:blockchoice:removeinnerouter}
\BlankLine
		$\RB \gets \RB \setminus \{ B \}$\;
		$\BSorted[t] \gets B$\;
		$tc \gets tc + \score{$B$}$ \tcp*[r]{update number of removed run-characters from tunneling}
\BlankLine
		\If(\tcp*[f]{update optimal block choice; see equations \eqref{eq:tunneling:gross-benefit} and \eqref{eq:tunneling:tax}})
		   {$\FuncSty{gross-benefit}-\FuncSty{tax} > \nbbest$} {
			$\tbest \gets t$\;
			$\nbbest \gets \FuncSty{gross-benefit}-\FuncSty{tax}$\;
		}		
	}
\caption{Greedy block choice}
\label{alg:blockchoice}
\end{algorithm}

The collision handling of Algorithm \ref{alg:blockchoice} can be done using a
collision graph, that is, a graph connecting colliding blocks whose intersecting
area is not overlaid by a third block. Line \ref{alg:blockchoice:reducescore} then
can be implemented using a graph traversal together with block information, while
Line \ref{alg:blockchoice:removeinnerouter} can be performed by removing the node
corresponding to block $B$ from the graph.

Implementing block scores is a bit complicated, as run-length-encoding is used.
Initially, the score of a Block $B$ gets to the sum of the
run-lengths of all runs $B$ points in minus the sum of all reduced run-lengths,
i.e.\, the sum of $\log( h ) - \log( h - h_{B} )$ for each run of height $h$ where 
$B$ points in. Score updates of outer collisions are approximated
by multiplying the score with the ratio by which the block width was shortened.
Score updates of inner collisions can be done by subtracting the
width-difference-ratio-multiplied score of $B$.%
\footnote{Let $\innU{B}$ be the inner colliding block of $B$.
	We set $\FuncSty{score}(\innU{B}) = \FuncSty{score}(\innU{B}) -
	\frac{w_{\innU{B}}}{w_B} \cdot \FuncSty{score}(B)$, which is motivated
	by the fact that $\log( h - h_B ) - \log( h - h_B - (h_B - h_{\innU{B}}) )
	= \log( h - h_B ) - \log( h - h_{\innU{B}} ) = \left( \log( h ) - \log( h - h_{\innU{B}} ) \right)
	- \left( \log( h ) - \log( h - h_B ) \right)$ for a single run in which
	$\innU{B}$ and $B$ collide.}%

By using a heap, Algorithm \ref{alg:blockchoice} can be shown to run in
$O(n \log |\RB|)$ time (Appendix \ref{sec:algblockchoicetime}).
Unfortunately, we have to mention that the greedy strategy is not always optimal: think of three blocks
whose colliding picture forms a shape like a big ``H''. If the middle block has
a score bigger than that of the outer blocks (but close enough), for $t=1$, the middle
block is the optimal choice, while for $t=2$ the outer blocks would be preferred,
what is not covered by the algorithm. These situations however should not
occur often in practice, so we neglected them.

\newcommand{\BWZ}{\texttt{bwz} }
\newcommand{\TBWZ}{\texttt{tbwz} }
\newcommand{\BCM}{\texttt{bcm} }
\newcommand{\TBCM}{\texttt{tbcm} }
\newcommand{\WT}{\texttt{wt} }
\newcommand{\TWT}{\texttt{twt} }
\newcommand{\XZ}{\texttt{xz} }
\newcommand{\ZPAQ}{\texttt{zpaq} }

\pgfplotstableread{
file bcm-bps bcm-decode-speed bcm-decode-membps bcm-encode-speed bcm-encode-membps bwz-bps bwz-decode-speed bwz-decode-membps bwz-encode-speed bwz-encode-membps bzip2-bps bzip2-decode-speed bzip2-decode-membps bzip2-encode-speed bzip2-encode-membps gzip-bps gzip-decode-speed gzip-decode-membps gzip-encode-speed gzip-encode-membps tbcm-bps tbcm-decode-speed tbcm-decode-membps tbcm-encode-speed tbcm-encode-membps tbwz-bps tbwz-decode-speed tbwz-decode-membps tbwz-encode-speed tbwz-encode-membps twt-bps twt-decode-speed twt-decode-membps twt-encode-speed twt-encode-membps wt-bps wt-decode-speed wt-decode-membps wt-encode-speed wt-encode-membps xz-extreme-bps xz-extreme-decode-speed xz-extreme-decode-membps xz-extreme-encode-speed xz-extreme-encode-membps xz-bps xz-decode-speed xz-decode-membps xz-encode-speed xz-encode-membps zpaq-bps zpaq-decode-speed zpaq-decode-membps zpaq-encode-speed zpaq-encode-membps bps-best decode-speed-best decode-membps-best encode-speed-best encode-membps-best
dickens 1.761 5.71 42.18 6.51 42.20 2.113 4.45 42.08 9.16 42.27 2.197 34.59 3.95 13.11 6.17 3.036 190.59 2.85 18.30 2.83 1.752 5.19 43.71 4.35 99.83 2.109 4.17 43.53 5.45 98.33 2.677 3.84 44.60 6.39 106.53 2.700 3.99 43.03 12.44 43.27 2.222 80.33 10.84 1.61 128.95 2.222 80.33 8.62 1.64 78.58 1.775 0.81 84.34 0.83 86.48 tbcm gzip gzip gzip gzip
mozilla 2.495 5.66 40.45 7.30 40.47 2.939 3.02 40.41 8.59 40.48 2.798 34.61 0.77 13.83 1.36 2.975 157.06 0.56 26.10 0.55 2.450 5.78 39.09 4.42 88.50 2.915 3.07 39.07 4.86 88.59 3.732 3.39 39.29 6.14 98.57 3.903 3.19 40.65 14.36 40.69 2.089 54.21 8.69 1.94 83.73 2.109 54.82 1.71 2.41 15.65 2.120 0.72 17.81 0.74 17.84 xz-extreme gzip gzip gzip gzip
mr 1.699 5.33 42.13 7.36 42.38 2.083 3.35 42.20 9.89 42.39 1.958 45.06 4.01 18.97 6.37 2.961 186.44 2.87 20.18 2.83 1.699 5.49 44.30 6.25 84.51 2.083 3.33 44.24 8.05 84.45 2.349 4.46 45.35 10.67 93.65 2.345 4.42 43.37 14.38 43.43 2.208 59.06 10.07 1.59 130.33 2.206 55.60 8.86 2.43 80.18 1.795 0.82 88.22 0.83 88.30 bcm gzip gzip gzip gzip
nci 0.292 6.24 40.62 7.82 40.69 0.339 10.45 40.67 17.19 40.70 0.432 58.07 1.21 7.74 1.93 0.762 288.27 0.84 72.56 0.87 0.275 6.55 35.92 5.94 40.74 0.328 9.60 35.94 9.03 40.70 0.542 7.84 36.23 9.27 41.05 0.608 7.65 40.91 18.27 41.07 0.345 244.26 8.65 1.17 89.40 0.424 226.94 2.61 3.33 23.88 0.362 0.93 27.02 0.97 27.17 tbcm gzip gzip gzip gzip
ooffice 3.306 5.98 43.47 7.51 43.60 3.821 2.36 43.36 7.60 43.62 3.722 26.54 6.56 13.30 10.90 4.027 143.10 4.55 20.87 4.54 3.265 5.69 43.58 4.37 115.95 3.797 2.39 43.05 4.54 115.87 5.138 2.93 45.31 6.23 131.11 5.332 2.84 45.17 14.27 45.58 3.156 41.61 13.26 2.60 167.17 3.155 38.85 12.29 2.66 99.56 2.931 0.70 137.52 0.70 137.67 zpaq gzip gzip gzip gzip
osdb 1.784 5.31 42.25 7.28 42.20 2.250 3.58 42.03 10.11 42.31 2.223 35.49 3.98 13.52 6.26 2.966 157.67 2.80 31.95 2.90 1.770 7.17 30.67 7.17 80.51 2.238 3.90 30.68 8.21 80.25 2.510 4.38 31.90 10.79 88.53 2.952 3.09 43.30 14.33 43.40 2.256 59.74 11.44 2.13 131.14 2.260 56.24 8.77 2.28 79.29 1.886 0.76 87.27 0.77 85.76 tbcm gzip gzip gzip gzip
reymont 1.186 5.90 43.26 7.25 43.46 1.470 3.39 43.09 9.56 43.37 1.504 39.25 5.95 14.01 9.59 2.243 203.87 4.29 20.32 4.23 1.184 5.53 44.40 5.17 84.46 1.470 3.22 44.16 6.37 83.56 1.966 4.44 45.85 8.19 85.19 1.996 4.44 44.92 14.33 45.11 1.588 103.60 12.84 1.67 161.25 1.589 103.60 11.11 1.77 99.88 1.271 0.86 127.66 0.87 127.86 tbcm gzip gzip gzip gzip
samba 1.488 6.95 41.04 7.65 41.01 1.805 4.10 40.94 10.56 41.01 1.684 46.72 1.85 13.91 2.99 2.021 204.01 1.35 41.96 1.26 1.419 7.65 34.37 3.90 66.64 1.750 4.45 34.27 4.38 67.19 2.445 4.54 34.84 5.09 74.51 2.696 3.87 41.54 15.96 41.60 1.384 89.20 8.97 1.77 99.53 1.402 89.20 4.08 3.12 37.09 1.196 0.80 42.25 0.83 42.33 zpaq gzip gzip gzip gzip
sao 5.155 4.73 43.17 6.51 43.14 5.949 2.06 42.98 6.06 43.21 5.450 22.97 5.56 10.78 8.76 5.882 168.68 4.05 18.15 3.95 5.155 4.48 45.37 5.23 157.95 5.949 2.00 45.04 4.90 157.16 6.234 2.43 46.59 8.22 172.56 6.227 2.41 44.48 11.70 44.65 4.882 25.52 11.87 2.75 154.14 4.870 24.61 11.58 2.61 97.46 4.980 0.63 118.92 0.64 119.14 xz gzip gzip gzip gzip
webster 1.239 5.79 40.49 6.63 40.58 1.481 5.52 40.51 9.90 40.59 1.668 39.10 0.96 12.12 1.56 2.354 178.90 0.67 28.83 0.66 1.236 5.46 41.76 4.52 74.38 1.480 5.46 41.75 5.99 74.54 2.074 4.37 42.04 6.70 81.74 2.075 4.42 40.79 12.16 40.83 1.614 87.66 8.74 1.40 86.48 1.665 87.66 2.11 1.73 19.33 1.209 0.81 22.00 0.85 22.03 zpaq gzip gzip gzip gzip
xml 0.590 8.34 44.16 8.62 44.35 0.664 12.71 44.02 18.81 44.51 0.660 56.01 7.33 9.25 12.97 1.035 242.74 5.44 56.01 5.23 0.559 11.30 34.06 6.13 49.95 0.643 15.40 34.20 8.48 52.20 1.163 7.07 36.06 9.25 56.12 1.334 5.36 46.17 20.30 46.40 0.650 242.74 13.55 2.33 182.16 0.678 164.44 14.28 4.46 103.94 0.530 0.88 158.32 0.90 155.35 zpaq xz-extreme gzip gzip gzip
x-ray 3.452 5.21 42.50 6.90 42.74 4.244 2.43 42.38 7.47 42.72 3.824 28.76 4.74 14.40 7.55 5.699 113.82 3.32 25.98 3.33 3.452 4.89 44.73 5.76 120.75 4.244 2.40 44.67 6.30 120.65 4.473 2.87 46.09 9.38 133.20 4.470 2.88 43.78 12.41 43.98 4.239 26.84 10.87 2.56 141.17 4.238 26.84 10.41 2.53 92.62 3.560 0.63 101.80 0.66 101.89 bcm gzip gzip gzip gzip
sources 1.222 6.35 40.10 6.82 40.11 1.383 3.87 40.10 9.23 40.11 1.493 45.90 0.19 12.50 0.33 1.804 195.05 0.13 35.96 0.13 1.165 6.89 34.89 2.89 63.28 1.349 4.14 34.88 3.18 63.25 2.082 4.68 34.98 3.48 70.00 2.286 4.15 40.15 12.96 40.16 1.184 103.60 2.64 1.67 26.87 1.264 97.57 0.41 2.49 3.80 0.989 0.81 4.33 0.85 4.33 zpaq gzip gzip gzip gzip
pitches 2.664 6.16 40.39 6.73 40.41 2.838 4.03 40.38 8.93 40.41 2.858 33.67 0.72 10.54 1.24 2.422 156.14 0.50 42.22 0.50 2.420 6.92 31.59 2.47 95.18 2.658 4.57 31.57 2.70 94.87 3.624 4.23 31.78 2.88 102.92 4.441 3.42 40.58 12.79 40.60 1.980 55.98 8.39 1.91 82.33 2.080 52.66 1.58 2.85 14.36 1.963 0.76 16.35 0.75 16.38 zpaq gzip gzip gzip gzip
proteins 2.331 5.46 40.01 5.00 40.02 2.289 7.54 40.01 6.86 40.01 3.451 25.57 0.03 11.31 0.05 3.575 124.62 0.02 29.51 0.02 1.949 5.85 28.38 1.49 86.75 2.003 7.51 28.38 1.58 86.76 2.715 4.77 28.43 1.60 92.29 3.971 3.96 40.02 7.67 40.02 2.222 52.34 0.47 0.97 4.78 2.575 43.81 0.07 1.53 0.67 2.609 0.74 0.76 0.72 0.76 tbcm gzip gzip gzip gzip
dna 1.720 5.63 40.05 5.83 40.06 1.829 7.17 40.05 8.10 40.06 2.060 32.58 0.09 12.91 0.15 2.249 169.62 0.07 10.10 0.07 1.696 4.80 41.27 3.11 124.34 1.808 6.23 41.26 3.83 124.30 2.027 5.16 41.31 4.18 128.56 2.047 5.82 40.07 9.78 40.08 1.778 75.22 1.38 0.75 14.02 1.818 80.40 0.21 1.02 1.98 1.859 0.94 2.23 0.96 2.21 tbcm gzip gzip bzip2 gzip
english.1024MB 1.561 5.20 40.02 5.63 40.02 1.837 2.59 40.02 6.41 40.02 2.266 32.42 0.03 12.96 0.05 3.037 155.59 0.02 18.56 0.02 1.337 5.51 33.20 1.50 80.83 1.658 2.92 33.20 1.54 80.84 1.987 4.34 33.21 1.64 86.42 2.450 3.84 40.03 9.16 40.03 1.925 77.68 0.51 1.13 5.27 2.098 72.36 0.08 1.55 0.74 1.641 0.79 0.85 0.80 0.85 tbcm gzip gzip gzip gzip
dblp.xml 0.628 6.71 40.07 6.83 40.08 0.751 9.23 40.07 11.91 40.07 0.911 48.35 0.13 10.39 0.21 1.393 217.07 0.09 53.27 0.09 0.617 7.54 32.98 5.23 48.70 0.744 9.40 32.99 6.98 48.72 1.163 6.51 33.03 7.40 54.08 1.266 5.84 40.11 13.06 40.11 0.819 145.50 1.88 1.52 19.13 0.942 130.68 0.30 2.91 2.70 0.605 0.84 3.08 0.87 3.08 zpaq gzip gzip gzip gzip
Escherichia-Coli 0.796 5.91 40.20 6.33 40.22 0.776 8.98 40.18 10.28 40.20 2.158 32.26 0.35 12.83 0.57 2.312 173.05 0.24 9.69 0.25 0.583 7.29 17.09 2.04 48.19 0.586 8.74 17.08 2.19 48.18 0.918 7.38 17.22 2.22 50.05 1.488 5.78 40.28 11.01 40.31 0.368 261.48 4.95 0.92 50.28 0.719 167.65 0.77 1.21 7.11 1.941 0.94 8.01 0.94 8.06 xz-extreme xz-extreme gzip bzip2 gzip
cere 0.238 6.09 40.04 6.36 40.05 0.237 10.75 40.04 11.06 40.05 2.017 33.96 0.08 13.79 0.13 2.171 174.50 0.06 10.02 0.06 0.170 7.73 16.64 3.36 40.05 0.168 9.84 16.64 3.72 40.05 0.313 7.89 16.67 3.67 40.07 0.377 6.71 40.06 11.25 40.07 0.087 477.65 1.21 1.01 12.28 1.863 89.76 0.19 1.07 1.73 1.771 0.94 1.92 0.96 1.94 xz-extreme xz-extreme gzip bzip2 gzip
coreutils 0.229 5.73 40.10 6.76 40.12 0.232 7.08 40.11 12.13 40.11 1.281 47.62 0.19 12.24 0.34 1.961 188.06 0.14 35.20 0.13 0.163 10.72 11.59 3.29 40.12 0.172 10.49 11.58 3.47 40.11 0.379 8.93 11.64 3.52 40.16 0.690 4.61 40.15 12.92 40.17 0.144 415.65 2.71 2.19 27.60 0.934 130.42 0.42 3.41 3.90 0.618 0.84 4.44 0.87 4.45 xz-extreme xz-extreme gzip gzip gzip
einstein.de.txt 0.022 7.89 40.24 7.04 40.26 0.015 20.05 40.23 14.59 40.26 0.345 61.38 0.42 9.75 0.72 2.494 173.11 0.30 24.77 0.29 0.015 10.41 17.90 8.26 40.26 0.013 15.48 17.87 11.02 40.25 0.178 9.54 18.00 10.97 40.37 0.399 7.36 40.35 14.12 40.37 0.008 675.27 6.01 4.86 60.53 0.008 675.27 0.95 9.43 8.63 0.007 0.85 9.84 0.91 9.85 zpaq xz gzip gzip gzip
einstein.en.txt 0.015 6.78 40.04 6.65 40.05 0.007 14.36 40.04 13.14 40.04 0.413 60.17 0.08 9.47 0.14 2.806 158.64 0.06 24.39 0.06 0.010 7.98 24.53 7.18 40.05 0.006 12.65 24.53 10.31 40.05 0.233 8.19 24.56 10.26 40.07 0.394 7.31 40.06 12.78 40.07 0.005 654.86 1.19 4.77 12.11 0.005 636.18 0.18 9.01 1.71 0.004 0.87 1.95 0.90 1.95 zpaq xz-extreme gzip gzip gzip
influenza 0.119 7.33 40.13 6.76 40.15 0.120 15.50 40.13 12.64 40.15 0.526 55.48 0.25 8.80 0.44 0.897 254.10 0.17 24.80 0.18 0.112 7.05 36.72 4.41 40.15 0.116 12.62 36.71 5.99 40.16 0.297 7.84 36.79 6.01 40.23 0.314 8.32 40.20 12.62 40.22 0.082 459.92 3.60 1.76 36.60 0.117 408.96 0.56 4.34 5.17 0.351 0.94 5.88 0.96 5.81 xz-extreme xz-extreme gzip gzip gzip
kernel 0.136 5.49 40.09 6.75 40.09 0.130 7.95 40.08 12.71 40.09 1.738 41.68 0.15 13.75 0.24 2.160 178.14 0.10 32.11 0.10 0.094 10.84 9.29 4.43 40.09 0.095 11.84 9.29 4.70 40.09 0.223 9.50 9.33 4.74 40.13 0.517 4.49 40.13 12.85 40.13 0.064 511.45 2.16 2.28 21.96 0.201 350.94 0.34 2.98 3.10 0.058 0.86 3.53 0.89 3.54 zpaq xz-extreme gzip gzip gzip
para 0.315 6.09 40.04 6.26 40.05 0.308 10.52 40.04 10.66 40.05 2.091 33.17 0.09 13.38 0.14 2.244 172.66 0.06 9.68 0.06 0.211 8.15 12.56 3.75 40.05 0.209 9.89 12.56 4.07 40.05 0.424 8.12 12.60 4.01 40.08 0.490 6.47 40.07 10.91 40.08 0.113 425.99 1.30 1.00 13.20 1.925 85.26 0.20 1.03 1.86 1.854 0.94 2.09 0.96 2.09 xz-extreme xz-extreme gzip bzip2 gzip
world-leaders 0.126 5.72 40.46 8.57 40.49 0.121 9.40 40.43 22.05 40.50 0.555 75.79 0.81 20.63 1.33 1.428 222.84 0.61 50.27 0.58 0.105 7.18 24.42 6.17 40.52 0.106 9.86 24.38 8.27 40.53 0.287 7.36 24.66 8.30 40.73 0.396 5.90 40.69 22.27 40.74 0.088 492.22 8.75 2.19 84.48 0.103 492.22 1.89 5.86 17.05 0.093 0.92 19.45 0.94 19.46 xz-extreme xz gzip gzip gzip
}{\expresulttbl}
\pgfplotstableread{
file bcmzip-numberofblocks bcmzip-bwtconstructiontime bcmzip-inputsize bcmzip-encodingtime bcmzip-bwtencodingsize bwzip-numberofblocks bwzip-bwtconstructiontime bwzip-inputsize bwzip-encodingtime bwzip-bwtencodingsize tbcmzip-numberofblocks tbcmzip-blockcomputationtime tbcmzip-bwtconstructiontime tbcmzip-inputsize tbcmzip-blockchoicetime tbcmzip-tunneledblockcount tbcmzip-auxencodingsize tbcmzip-tbwtencodingsize tbcmzip-blockcount tbcmzip-expectedauxtaxsize tbcmzip-expectedtbwtgrossbenefitsize tbcmzip-encodingtime tbcmzip-tunnelingtime tbwzip-numberofblocks tbwzip-blockcomputationtime tbwzip-bwtconstructiontime tbwzip-inputsize tbwzip-blockchoicetime tbwzip-tunneledblockcount tbwzip-auxencodingsize tbwzip-tbwtencodingsize tbwzip-blockcount tbwzip-expectedauxtaxsize tbwzip-expectedtbwtgrossbenefitsize tbwzip-encodingtime tbwzip-tunnelingtime twtzip-numberofblocks twtzip-blockcomputationtime twtzip-bwtconstructiontime twtzip-inputsize twtzip-blockchoicetime twtzip-tunneledblockcount twtzip-auxencodingsize twtzip-tbwtencodingsize twtzip-blockcount twtzip-expectedauxtaxsize twtzip-expectedtbwtgrossbenefitsize twtzip-encodingtime twtzip-tunnelingtime wtzip-numberofblocks wtzip-bwtconstructiontime wtzip-inputsize wtzip-encodingtime wtzip-bwtencodingsize
dickens 1 622 10192446 808 2243740 1 628 10192446 429 2693077 1 494 621 10192446 147 1034 3268 2229704 93288 2641 10979 903 41 1 489 619 10192446 147 1034 4172 2683150 93288 2641 10979 461 41 1 496 625 10192446 148 1034 16766 3394146 93288 2641 10979 180 41 1 621 10192446 144 3439954
mozilla 1 2385 51220480 4255 15977017 1 2392 51220480 3294 18822650 1 2605 2381 51220480 1051 65536 110068 15579782 342715 128692 363611 4261 649 1 2591 2389 51220480 1054 65536 141792 18526835 342715 128692 363611 3321 650 1 2603 2383 51220480 1054 65536 179022 23719850 342715 128692 363611 1085 648 1 2383 51220480 1006 24993570
mr 1 512 9970564 752 2117647 1 508 9970564 426 2596965 1 155 509 9970564 26 34 218 2117521 18644 93 187 784 30 1 155 509 9970564 25 34 156 2596857 18644 93 187 445 30 1 155 509 9970564 26 34 6126 2922586 18644 93 187 161 30 1 506 9970564 146 2922794
nci 1 1568 33553445 2409 1227036 1 1565 33553445 260 1425554 1 800 1564 33553445 378 10532 19510 1135909 72835 22406 87737 2101 461 1 793 1567 33553445 372 10532 25453 1350268 72835 22406 87737 270 458 1 805 1574 33553445 376 10532 33030 2243458 72835 22406 87737 181 463 1 1567 33553445 166 2550146
ooffice 1 252 6152192 537 2543034 1 250 6152192 521 2939008 1 329 252 6152192 124 8192 15919 2495020 63245 17428 46466 569 56 1 328 249 6152192 123 8192 20629 2899911 63245 17428 46466 525 56 1 330 250 6152192 125 8192 27430 3924010 63245 17428 46466 169 56 1 252 6152192 150 4100610
osdb 1 510 10085684 811 2249281 1 510 10085684 438 2836862 1 148 507 10085684 27 326 832 2231503 19433 832 14609 606 41 1 147 510 10085684 27 326 992 2821196 19433 832 14609 438 41 1 148 509 10085684 27 326 6678 3158698 19433 832 14609 151 41 1 514 10085684 155 3721970
reymont 1 359 6627202 511 983171 1 361 6627202 287 1218219 1 221 357 6627202 68 513 1521 979803 49560 1310 2624 546 22 1 218 358 6627202 68 513 1934 1215988 49560 1310 2624 322 22 1 221 360 6627202 68 513 9094 1620266 49560 1310 2624 93 22 1 357 6627202 76 1654178
samba 1 950 21606400 1704 4019286 1 958 21606400 975 4876339 1 1482 958 21606400 703 32768 52997 3780007 158532 64346 256549 1489 522 1 1473 955 21606400 703 32768 69217 4658017 158532 64346 256549 947 521 1 1478 955 21606400 700 32768 86870 6517386 158532 64346 256549 324 519 1 954 21606400 306 7282746
sao 1 342 7251944 719 4673281 1 342 7251944 811 5392986 1 169 340 7251944 36 12 149 4673237 21753 34 37 750 22 1 167 342 7251944 35 12 78 5392929 21753 34 37 841 22 1 168 342 7251944 36 12 6014 5645186 21753 34 37 257 22 1 341 7251944 242 5645498
webster 1 2690 41458703 3217 6421306 1 2698 41458703 1235 7678044 1 1742 2683 41458703 599 4164 12499 6394497 308199 10282 22584 3487 165 1 1722 2687 41458703 592 4164 16144 7657985 308199 10282 22584 1340 164 1 1737 2686 41458703 600 4164 46582 10704194 308199 10282 22584 597 164 1 2691 41458703 483 10755690
xml 1 199 5345280 399 394620 1 197 5345280 76 443860 1 162 198 5345280 78 4097 6585 367459 25002 8045 26042 307 61 1 169 198 5345280 82 4097 8401 421268 25002 8045 26042 77 65 1 168 196 5345280 80 4097 13222 764314 25002 8045 26042 44 65 1 197 5345280 44 891778
x-ray 1 427 8474240 748 3657057 1 426 8474240 652 4495602 1 137 426 8474240 15 0 31 3657057 5031 0 0 786 28 1 137 423 8474240 15 0 12 4495602 5031 0 0 669 28 1 138 425 8474240 15 0 2638 4735514 5031 0 0 236 28 1 424 8474240 220 4735514
sources 1 12905 210866607 16306 32231063 1 12820 210866607 8755 36474659 1 22357 12803 210866607 11171 342351 603646 30126053 2197741 728332 2022640 15176 6723 1 22227 12801 210866607 11147 342351 793154 34788769 2197741 728332 2022640 8715 6705 1 22334 12812 210866607 11180 342351 938070 53958490 2197741 728332 2022640 2818 6715 1 12803 210866607 2532 60256274
pitches 1 3045 55832855 4714 18592319 1 3051 55832855 2916 19809376 1 7433 3048 55832855 3532 262144 371750 16524628 769960 458752 2040071 4045 2530 1 7392 3046 55832855 3539 262144 485620 18069209 769960 458752 2040071 2685 2524 1 7411 3051 55832855 3534 262144 625694 24671398 769960 458752 2040071 1035 2523 1 3049 55832855 1019 30996822
proteins 1 124633 1184051855 96312 345114688 1 124135 1184051855 36501 338928139 1 303876 126056 1184051855 124228 2861875 4428810 284123012 10297851 6088478 52601999 73257 106677 1 290177 123959 1184051855 116726 2861875 5889558 290656854 10297851 6088478 52601999 32214 102210 1 301718 126222 1184051855 121699 2861875 7171886 394714694 10297851 6088478 52601999 16705 107136 1 124092 1184051855 16313 587751038
dna 1 35118 403927746 30795 86891881 1 34796 403927746 12632 92386548 1 33823 34760 403927746 12815 84582 257882 85392496 7322290 216091 1502069 36409 4674 1 33379 34653 403927746 12773 84582 335628 90955875 7322290 216091 1502069 14081 4664 1 34069 35156 403927746 13227 84582 893942 101467970 7322290 216091 1502069 6361 4818 1 34763 403927746 4479 103396634
english.1024MB 1 96688 1073741824 84433 209547301 1 97624 1073741824 61857 246663166 1 278092 96644 1073741824 116259 271760 702582 178767268 7206829 671100 28463995 75668 111958 1 277664 96921 1073741824 116311 271760 908353 221628558 7206829 671100 28463995 55730 111844 1 288299 97605 1073741824 121438 271760 1574870 265131398 7206829 671100 28463995 14876 116574 1 97813 1073741824 13648 328848894
dblp.xml 1 18962 296135874 22215 23251073 1 19190 296135874 4524 27831941 1 9646 19131 296135874 4303 131073 229412 22618854 1298281 278850 681389 18721 2587 1 9487 18953 296135874 4148 131073 299476 27266442 1298281 278850 681389 4736 2548 1 9534 18942 296135874 4167 131073 387878 42665726 1298281 278850 681389 2534 2547 1 18960 296135874 2366 46867670
Escherichia-Coli 1 8819 112689515 8115 11226044 1 8821 112689515 1591 10943462 1 21152 8819 112689515 8807 262144 406822 7806636 894789 514769 3412475 4011 9820 1 21150 8826 112689515 8789 262144 541679 7716932 894789 514769 3412475 1366 9804 1 20552 8753 112689515 8421 262144 661614 12279542 894789 514769 3412475 596 9502 1 8765 112689515 894 20971638
cere 1 36869 461286644 32136 13766371 1 37153 461286644 2647 13690609 1 40277 36831 461286644 16533 182152 318635 9504297 778899 387518 4112529 13658 22295 1 40337 36909 461286644 16536 182152 427168 9274744 778899 387518 4112529 1839 22321 1 40363 36851 461286644 16553 182152 485478 17612670 778899 387518 4112529 1380 22310 1 36874 461286644 2105 21742094
coreutils 1 13800 205281778 14947 5901043 1 13803 205281778 2255 5974458 1 19986 13796 205281778 9210 42144 77022 4126127 282273 95264 2052339 4478 12055 1 19827 13893 205281778 9130 42144 98882 4325422 282273 95264 2052339 1316 11916 1 20180 13851 205281778 9370 42144 123350 9605498 282273 95264 2052339 583 12152 1 13891 205281778 1202 17725570
einstein.de.txt 1 5875 92758441 6684 266412 1 5904 92758441 198 185439 1 931 5842 92758441 270 2063 3376 177386 9370 4051 38070 2854 765 1 931 5928 92758441 269 2063 4458 147152 9370 4051 38070 132 763 1 922 5870 92758441 263 2063 8070 2056894 9370 4051 38070 187 756 1 5902 92758441 359 4635238
einstein.en.txt 1 32975 467626544 33727 897148 1 33256 467626544 875 444842 1 3930 32923 467626544 1416 4100 6929 610861 19440 8051 75797 19897 3627 1 3963 33330 467626544 1436 4100 9188 371244 19440 8051 75797 755 3614 1 3992 32924 467626544 1425 4100 13550 13646158 19440 8051 75797 1229 3610 1 32999 467626544 1780 23047110
influenza 1 10987 154808555 10740 2309584 1 10954 154808555 603 2334817 1 6244 10928 154808555 3392 32768 62033 2110879 187503 64346 182497 9643 2961 1 6333 11026 154808555 3524 32768 82623 2170652 187503 64346 182497 702 3051 1 6279 10983 154808555 3385 32768 94422 5660430 187503 64346 182497 667 2965 1 10953 154808555 611 6088854
kernel 1 17692 257961616 18628 4394953 1 17759 257961616 1691 4219289 1 17833 17689 257961616 6114 22344 45819 3002745 163797 50507 1464293 4368 9887 1 17851 17749 257961616 6137 22344 59569 3014528 163797 50507 1464293 895 9857 1 17632 17654 257961616 5994 22344 71806 7125042 163797 50507 1464293 507 9773 1 17640 257961616 1273 16693866
para 1 35242 429265758 29977 16913650 1 35230 429265758 2936 16556030 1 33729 35195 429265758 14153 349289 550558 10778405 1238462 685894 6425006 10082 15393 1 33760 35219 429265758 14180 349289 734233 10480315 1238462 685894 6425006 1908 15389 1 33684 35199 429265758 14155 349289 885510 21883278 1238462 685894 6425006 1248 15329 1 35201 429265758 2181 26295486
world-leaders 1 1781 46968181 3411 743036 1 1781 46968181 222 713431 1 1382 1781 46968181 791 16384 23340 595101 55339 28672 131073 2004 1231 1 1383 1798 46968181 780 16384 30373 593570 55339 28672 131073 183 1235 1 1383 1788 46968181 793 16384 41038 1645902 55339 28672 131073 155 1234 1 1794 46968181 208 2330054
}{\bwinfotbl}

\pgfplotstableset{
	create on use/tbwz-comp-improve-percentage/.style={
		create col/expr={100 - \thisrow{tbwz-bps}*100/\thisrow{bwz-bps}}
	},
	create on use/tbwz-encspeed-increase-percentage/.style={
		create col/expr={\thisrow{bwz-encode-speed}*100/\thisrow{tbwz-encode-speed} - 100}
	},
	create on use/tbwz-decspeed-increase-percentage/.style={
		create col/expr={\thisrow{bwz-decode-speed}*100/\thisrow{tbwz-decode-speed} - 100}
	},
	create on use/tbwz-encmempeak-increase-percentage/.style={
		create col/expr={\thisrow{tbwz-encode-membps}*100/\thisrow{bwz-encode-membps} - 100}
	},
	create on use/tbwz-decmempeak-increase-percentage/.style={
		create col/expr={\thisrow{tbwz-decode-membps}*100/\thisrow{bwz-decode-membps} - 100}
	},
	create on use/tbcm-comp-improve-percentage/.style={
		create col/expr={100 - \thisrow{tbcm-bps}*100/\thisrow{bcm-bps}}
	},
	create on use/tbcm-encspeed-increase-percentage/.style={
		create col/expr={\thisrow{bcm-encode-speed}*100/\thisrow{tbcm-encode-speed} - 100}
	},
	create on use/tbcm-decspeed-increase-percentage/.style={
		create col/expr={\thisrow{bcm-decode-speed}*100/\thisrow{tbcm-decode-speed} - 100}
	},
	create on use/tbcm-encmempeak-increase-percentage/.style={
		create col/expr={\thisrow{tbcm-encode-membps}*100/\thisrow{bcm-encode-membps} - 100}
	},
	create on use/tbcm-decmempeak-increase-percentage/.style={
		create col/expr={\thisrow{tbcm-decode-membps}*100/\thisrow{bcm-decode-membps} - 100}
	},
	create on use/twt-comp-improve-percentage/.style={
		create col/expr={100 - \thisrow{twt-bps}*100/\thisrow{wt-bps}}
	},
	create on use/twt-encspeed-increase-percentage/.style={
		create col/expr={\thisrow{wt-encode-speed}*100/\thisrow{twt-encode-speed} - 100}
	},
	create on use/twt-decspeed-increase-percentage/.style={
		create col/expr={\thisrow{wt-decode-speed}*100/\thisrow{twt-decode-speed} - 100}
	},
	create on use/twt-encmempeak-increase-percentage/.style={
		create col/expr={\thisrow{twt-encode-membps}*100/\thisrow{wt-encode-membps} - 100}
	},
	create on use/twt-decmempeak-increase-percentage/.style={
		create col/expr={\thisrow{twt-decode-membps}*100/\thisrow{wt-decode-membps} - 100}
	},
	create on use/bcm-model-fit/.style={
		create col/expr={
			max(0,min(
			(1 + \thisrow{tbcmzip-auxencodingsize} / (\thisrow{bcmzip-bwtencodingsize} - \thisrow{tbcmzip-tbwtencodingsize} - \thisrow{tbcmzip-auxencodingsize}))
			,
			(1 + \thisrow{tbcmzip-expectedauxtaxsize} / (\thisrow{tbcmzip-expectedtbwtgrossbenefitsize} - \thisrow{tbcmzip-expectedauxtaxsize}))
			)) / max(
			(1 + \thisrow{tbcmzip-auxencodingsize} / (\thisrow{bcmzip-bwtencodingsize} - \thisrow{tbcmzip-tbwtencodingsize} - \thisrow{tbcmzip-auxencodingsize}))
			,
			(1 + \thisrow{tbcmzip-expectedauxtaxsize} / (\thisrow{tbcmzip-expectedtbwtgrossbenefitsize} - \thisrow{tbcmzip-expectedauxtaxsize}))
			)
			* 100
		}
	},
	create on use/bwz-model-fit/.style={
		create col/expr={
			max(0,min(
			(1 + \thisrow{tbwzip-auxencodingsize} / (\thisrow{bwzip-bwtencodingsize} - \thisrow{tbwzip-tbwtencodingsize} - \thisrow{tbwzip-auxencodingsize}))
			,
			(1 + \thisrow{tbwzip-expectedauxtaxsize} / (\thisrow{tbwzip-expectedtbwtgrossbenefitsize} - \thisrow{tbwzip-expectedauxtaxsize}))
			)) / max(
			(1 + \thisrow{tbwzip-auxencodingsize} / (\thisrow{bwzip-bwtencodingsize} - \thisrow{tbwzip-tbwtencodingsize} - \thisrow{tbwzip-auxencodingsize}))
			,
			(1 + \thisrow{tbwzip-expectedauxtaxsize} / (\thisrow{tbwzip-expectedtbwtgrossbenefitsize} - \thisrow{tbwzip-expectedauxtaxsize}))
			)
			* 100
		}
	},
	create on use/wt-model-fit/.style={
		create col/expr={
			max(0,min(
			(1 + \thisrow{twtzip-auxencodingsize} / (\thisrow{wtzip-bwtencodingsize} - \thisrow{twtzip-tbwtencodingsize} - \thisrow{twtzip-auxencodingsize}))
			,
			(1 + \thisrow{twtzip-expectedauxtaxsize} / (\thisrow{twtzip-expectedtbwtgrossbenefitsize} - \thisrow{twtzip-expectedauxtaxsize}))
			)) / max(
			(1 + \thisrow{twtzip-auxencodingsize} / (\thisrow{wtzip-bwtencodingsize} - \thisrow{twtzip-tbwtencodingsize} - \thisrow{twtzip-auxencodingsize}))
			,
			(1 + \thisrow{twtzip-expectedauxtaxsize} / (\thisrow{twtzip-expectedtbwtgrossbenefitsize} - \thisrow{twtzip-expectedauxtaxsize}))
			)
			* 100
		}
	}
}


\section{Experimental Results} \label{sec:expresults}
This section contains experimental results showing the effectiveness of our
new technique. We applied our technique as described in the last section to
three different $\BWT$ compressors:
\begin{itemize}
\item	\BWZ: \texttt{bzip2} \cite{bzip2-compressor} without memory limitations.
\item	\BCM: one of the strongest open-source-available $\BWT$ compressors \cite{bcm-compressor,enwik8benchmark}.
\item	\WT: $\BWT$ encoded in a huffman-shaped wavelet tree%
\footnote{Note that this data structure is \textbf{not capable} of text indexing.
	By permuting the bits in $\cntL$ according to the permutation induced by stably sorting column $\LCol$,
	the generalized LF-mapping can be computed using
	$\select_{\cntF}(1, \rank_{\cntL}(1,\C_{\LCol}[\LCol[i]] + \rank_{\LCol}(\LCol[i], i)))$,
	allowing backward steps using wavelet trees. Unfortunately, 
	more technical problems must be solved,
	outreaching the scope of this paper.}~
	using hybrid bitvectors \cite{FOS:GROS:GUP:BIT:2006,KAE:KEM:PUG:2014}
	(good compression for miscellaneous data \cite{KAE:KEM:PUG:2014})
	provided by the sdsl-lite library \cite{sdsl-lite-library}.
\end{itemize}

Our test data comes from three different text corpora, namely
12 medium-sized files from the Silesia Corpus \cite{silesia-corpus} (6 -- 49 MB),
6 bigger files from the Pizza \& Chili Corpus \cite{pizzachili-corpus} (54 -- 1130 MB),
as well as 9 repetitive files from the Repetitive Corpus \cite{repetitive-corpus}
(45 -- 446 MB). The full benchmark and all results are available at \cite{tbwz-compressor}.

\begin{figure}[h]
\pgfplotscreateplotcyclelist{bwboxplot}{%
{draw=black,fill=blue!5,mark=diamond*,mark options={fill=blue,fill opacity=.1,scale=1.5}},%
{draw=black,fill=blue!5,mark=diamond*,mark options={fill=blue,fill opacity=.1,scale=1.5}}%
}
\pgfplotsset{
	boxplot/draw/average/.code={%
	\draw[/pgfplots/boxplot/every average/.try]
		(boxplot box cs:\pgfplotsboxplotvalue{average},0.05)
		--
		(boxplot box cs:\pgfplotsboxplotvalue{average},0.95)
		;
	},
	boxplot/whisker range={1.5},
	boxplot/average={auto},
	boxplot/every box/.style={draw,fill},
	boxplot/every whisker/.style={},
	boxplot/every median/.style={thick},
	boxplot/every average/.style={semithick,dashed,red},
	cycle list name=bwboxplot,
	y=8ex,
	clip=false,
	xmajorgrids=true,
	xtick align=outside,tickwidth={.075cm},
	xticklabel={$\phantom{\scriptstyle\%} \pgfmathprintnumber{\tick} \scriptstyle\%$}
}
\tiny
	\begin{tikzpicture}
	\begin{axis}[
		title={\footnotesize Encoding Size Decrease},
		ytick={1,2,3},yticklabels={\footnotesize\WT,\footnotesize\BCM,\footnotesize\BWZ},
		separate axis lines,
		axis x line=bottom,axis y line=left,y axis line style={arrows=-},
		enlarge x limits=auto,enlarge y limits=auto,
		width={.7\textwidth}]
	\addplot+[boxplot] table[y=twt-comp-improve-percentage] \expresulttbl
		node[right,/pgfplots/boxplot/every average/.try] at
		(boxplot box cs: \boxplotvalue{average},0.5)
		{$\pgfmathprintnumber{\boxplotvalue{average}}$};

	\addplot+[boxplot] table[y=tbcm-comp-improve-percentage] \expresulttbl
		node[right,/pgfplots/boxplot/every average/.try] at
		(boxplot box cs: \boxplotvalue{average},0.5)
		{$\pgfmathprintnumber{\boxplotvalue{average}}$};

	\addplot+[boxplot] table[y=tbwz-comp-improve-percentage] \expresulttbl
		node[right,/pgfplots/boxplot/every average/.try] at
		(boxplot box cs: \boxplotvalue{average},0.5)
		{$\pgfmathprintnumber{\boxplotvalue{average}}$};
	\end{axis}
	
	\begin{axis}[
		title={\footnotesize Model Fit},
		ytick={1,2,3},
		separate axis lines,
		axis x line=bottom,axis y line=none,x axis line style={arrows=-},
		enlarge x limits=auto,
		xshift={.625\textwidth},width={.4\textwidth}]
	\addplot+[boxplot/average={},boxplot] table[y=wt-model-fit] \bwinfotbl;

	\addplot+[boxplot/average={},boxplot] table[y=bcm-model-fit] \bwinfotbl;

	\addplot+[boxplot/average={},boxplot] table[y=bwz-model-fit] \bwinfotbl;
	\end{axis}
	\end{tikzpicture}
\caption{Compression improvements and model fit of tunneling displayed as Tukey boxplots.
	The boxplots contain the whole tested data set. Boxes consist of
	lower quantile, median, upper quartile and average (red dashed line), whisker range
	is given by $1.5$ times the interquartile range, outliers are shown as diamond markers.
	Compression improvements use the untunneled versions as baseline,
	model fits are given by the min-max distance of the gross-net benefit ratios of
	theoretical model and compressor.}
\label{fig:bwz-tbwz-comparison}
\end{figure}
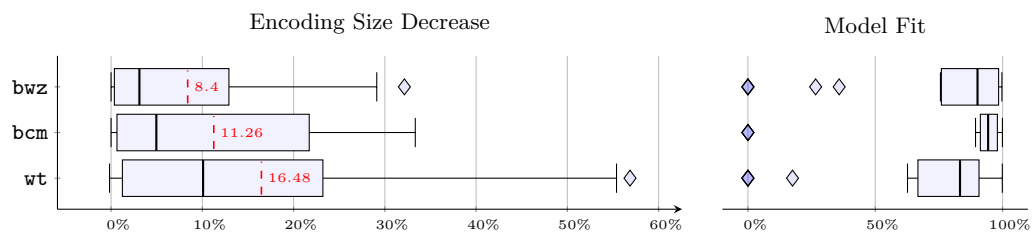

Beside of compression improvements, it is important to see if tunneling
exploits its full potential not only in theory (as shown in the heuristics
 and approximations from last section) but also in practice. Therefore, we
measured how well the theoretical model fits to the respective compressor
by comparing the gross-net benefit ratios of model and compressor: as the
efficiency of a compressor can be seen as some constant which is canceled when
dividing two benefits from the same source, the gross-net benefit ratio should
mostly be independent from the efficiency of a compressor, making it nicely
comparable. Figure \ref{fig:bwz-tbwz-comparison} shows compression improvements
and the model fit as min-max distance $\frac{\min\{x,y\}}{\max\{x,y\}}$
of both ratios.

As the box plots show, the compression improvements of tunneling are significant
and the theoretical model fits quite well. However, for half of the test files the encoding
size decrease lies below $4 - 11 \%$--not very surprisingly, this half typically consists
of small to medium-sized files where the normal $\BWT$-based compressors work very good already.
Compression improvements for the upper half of the data however are significant and range from
$4 - 11 \%$ encoding size decrease up to $33 - 57 \%$ decrease. Also, it seems that the better the
compressor works (in terms of compression rate), the better the model fits.

The outliers in the model fit can
be ignored for two reasons: first, as the figure shows, tunneling never worsens
compression by a serious amount; second, the potential of tunneling in all those
files (fraction of net-benefit of theoretical model and the size of
the \BWZ encoding) is below $0.3 \%$. The model fit itself however
shouldn't be overestimated, as we tested tunneling with different
models, always getting a quite similar compression result.


\newcommand{\findmin}[3]{
    \pgfplotstablevertcat{\findmintable}{#1}
    \pgfplotstablecreatecol[
      create col/expr={%
    \pgfplotstablerow
    }]{rownumber}\findmintable
    \pgfplotstablesort[sort key={#2},sort cmp={float <}]{\sorted}{\findmintable}%
    \pgfplotstablegetelem{0}{rownumber}\of{\sorted}%
    \pgfmathtruncatemacro#3{\pgfplotsretval}
    \pgfplotstableclear{\findmintable}
}
\pgfplotstableset{
    highlight col min/.code 2 args={
        \findmin{#1}{#2}{\minval}
        \edef\setstyles{\noexpand\pgfplotstableset{
                every row \minval\noexpand\space column #2/.style={
                    postproc cell content/.append style={
                        /pgfplots/table/@cell content/.add={$\noexpand\bf}{$}
                    },
                }
            }
        }\setstyles
    }
}

\begin{table}[ht]
\centering
\pgfplotstabletranspose*[colnames from=file,columns={file,
	bwz-bps,tbwz-bps,
	bcm-bps,tbcm-bps,
	wt-bps,twt-bps,
	xz-extreme-bps,
	zpaq-bps}]{\tmptable}{\expresulttbl}
\renewcommand{\arraystretch}{1.75}
\renewcommand{\tabcolsep}{.75em}
\tiny
\pgfplotstabletypeset[columns={colnames,nci,samba,webster,proteins,dna,english.1024MB,coreutils,para,world-leaders}
                     ,every head row/.style={
                      	before row={   \multicolumn{1}{c}{\footnotesize Compressor}
                      	             & \multicolumn{3}{c}{\footnotesize Silesia Corpus}
                      	             & \multicolumn{3}{c}{\footnotesize Pizza \& Chili Corpus}
                     	             & \multicolumn{3}{c}{\footnotesize Repetitive Corpus} \\ \hline
                      	},
                      	after row={\hline}
                      }
                     ,columns/colnames/.style={string type,column name={},column type={l},
                                               string replace={bwz-bps}{\BWZ},
                                               string replace={tbwz-bps}{\BWZ\texttt{-tunneled}},
                                               string replace={bcm-bps}{\BCM},
                                               string replace={tbcm-bps}{\BCM\texttt{-tunneled}},
                                               string replace={wt-bps}{\WT},
                                               string replace={twt-bps}{\WT\texttt{-tunneled}},
                                               string replace={xz-extreme-bps}{\XZ\texttt{-9e -M 100\%} \hfill\cite{xz-compressor}},
                                               string replace={zpaq-bps}{\ZPAQ\texttt{isc} \hfill\cite{zpaq-compressor}}}
                     ,highlight col min={\tmptable}{nci}
                     ,highlight col min={\tmptable}{samba}
                     ,highlight col min={\tmptable}{webster}
                     ,highlight col min={\tmptable}{proteins}
                     ,highlight col min={\tmptable}{dna}
                     ,highlight col min={\tmptable}{english.1024MB}
                     ,highlight col min={\tmptable}{coreutils}
                     ,highlight col min={\tmptable}{para}
                     ,highlight col min={\tmptable}{world-leaders}
                     ,columns/nci/.style={          column name={\tiny\rotatebox{90}{\parbox[t]{4.75em}{nci\newline(32 MB)}}},
                                                    column type={|c}}
                     ,columns/samba/.style={        column name={\tiny\rotatebox{90}{\parbox[t]{4.75em}{samba\newline(21 MB)}}}}
                     ,columns/webster/.style={      column name={\tiny\rotatebox{90}{\parbox[t]{4.75em}{webster\newline(40 MB)}}}}
                     ,columns/proteins/.style={     column name={\tiny\rotatebox{90}{\parbox[t]{4.75em}{proteins\newline(1130 MB)}}},
                                                    column type={|c}}
                     ,columns/dna/.style={          column name={\tiny\rotatebox{90}{\parbox[t]{4.75em}{dna\newline(386 MB)}}}}
                     ,columns/english.1024MB/.style={      column name={\tiny\rotatebox{90}{\parbox[t]{4.75em}{english\newline(1024 MB)}}}}
                     ,columns/coreutils/.style={    column name={\tiny\rotatebox{90}{\parbox[t]{4.75em}{coreutils\newline(196 MB)}}},
                                                    column type={|c}}
                     ,columns/para/.style={         column name={\tiny\rotatebox{90}{\parbox[t]{4.75em}{para\newline(410 MB)}}}}
                     ,columns/world-leaders/.style={column name={\tiny\rotatebox{90}{\parbox[t]{4.75em}{world-leaders\newline(45 MB)} }}}
                     ,fixed,fixed zerofill,precision=2
	]{\tmptable}
\caption{Compression comparison of different compressors on a selection of the used test data.
	All values are shown in bits per symbol, that is, original file size times bits per symbol
	gives the compression size. Best compression results are marked bold.}
\label{tbl:compressor-comparison}
\end{table}

Table \ref{tbl:compressor-comparison} shows a comparison of our technique with
compressors following different paradigms: \XZ \cite{xz-compressor} is a very effective
 Lempel-Ziv compressor similar to \texttt{7zip} \cite{7zip-compressor}, while 
\ZPAQ \cite{zpaq-compressor} uses context-mixing. As the table shows, the tunneled version
of \BCM performs best among all shown $\BWT$ compressors, \XZ remains the best
choice for repetitive data. Beside of pure compression, we want to note that tunneling
has its price: the time and space requirements for encoding roughly double,
while decoding time and space requirements are reduced by a small amount.


\section{Conclusion} \label{sec:conclusion}
As we have seen in the last Section, compression gains due to our technique are a
significant improvement to $\BWT$-based compression. The technique however is in
early stage of development, and therefore has some outstanding problems like making
a good and economic block choice. A solution
might be to use heuristics for similar problems in the LCP array
\cite{DIN:FIS:KLO:LOE:SAD:2017,MAU:BEL:OHL:2017}, our presented approach is a nice
baseline but too complicated and resource-expensive. It also would be nice to get
rid of the restriction of run-based blocks; Section 
\ref{sec:practicalimplementation:blockcomputation} indicates that this
is possible, but collisions complicate the situation.
In our opinion, the ``big deal'' will be to build a compressed FM-index
\cite{FER:MAN:2005} with full functionality; the footnote on page \pageref{sec:expresults}
clearly indicates that this should be possible, although correct pattern counting
might be a bit tricky. Thinking of a text index with half of the size of the currently
best implementations seems utopian, but this paper shows that this goal
should be achievable, giving a lot of motivation for further research on the topic.

\bibliographystyle{plainurl}
\bibliography{bibliography}

\clearpage
\appendix

\section{Proof of Lemma \ref{lem:rankpreserve}} \label{sec:proofrankpreserve}
\newcommand{\LColt}{\widetilde{\LCol}}
\newcommand{\LFt}{\widetilde{\LF}}
\newcommand{\cntLt}{\widetilde{\cntL}}
\newcommand{\cntFt}{\widetilde{\cntF}}
\newcommand{\LFarg}[1]{\mathsf{\LF_{\text{arg}}(#1)}}

Let $\LCol$ be the Burrows-Wheeler-Transformation of a string of length $n$ with LF-mapping $\LF$ and 
let $\cntL$ and $\cntF$ be two bitvectors of size $n$. Then following properties hold
for the generalized LF-mapping $\LF_{\cntF}^{\cntL}$:
\begin{enumerate}
\setlength\itemsep{1ex}
\item	Let $\cntF[i] = \cntL[i] = 1$ for all $i \in [1,n]$.
	Then, $\LF_{\cntF}^{\cntL}$ is identical to the normal LF-mapping $\LF$.
\begin{proof}
\begin{subequations}
\begin{align*}
\LF_{\cntF}^{\cntL}[i] & = \select_{\cntF}\left(1, \sum_{k=1}^{n} \cntL[k] \cdot \indicator{\LCol[k] < \LCol[i]} +
                            \sum_{k=1}^{i} \cntL[k] \cdot \indicator{\LCol[k] = \LCol[i]} \right) \\
                       & = \C_{\LCol}[\LCol[i]] + \rank_{\LCol}( \LCol[i], i ) = \LF[i]
\end{align*}
\end{subequations}
\end{proof}

\item	Let $\cntF$ and $\cntL$ be two bitvectors such that $\LF_{\cntF}^{\cntL}[j] = \LF[j]$ for all
	$j$ with $\cntL[j] = 1$. Let $i$ be an integer with $\cntL[i] = \cntF\left[\LF[i]\right] = 1$.
	Crossing out position $i$ in $\LCol$ and position $\LF[i]$ in $\FCol$
	(setting $\cntL[i] = \cntF\left[\LF[i]\right] = 0$) does not change the mapping: Define
	\[
	\widetilde{\cntL}[j] \coloneqq \cntL[j] \cdot \indicator{j \neq i}
	\text{\quad and \quad}
	\widetilde{\cntF}[j] \coloneqq \cntF[j] \cdot \indicator{j \neq \LF[i]}
	\]
	Then, $\LF_{\widetilde{\cntF}}^{\widetilde{\cntL}}[j] = \LF[j]$ 
	for all $j$ with $\widetilde{\cntL}[j] = 1$.
\label{proof:rankpreserve:clearingvalues}
\begin{proof}
\begin{subequations}
\begin{align}
&\LF_{\cntFt}^{\cntLt}[j] 
   = \select_{\cntFt}\left( 1, \sum_{k=1}^{n} \cntLt[k] \cdot \indicator{\LCol[k] < \LCol[j]} +
     \sum_{k=1}^{j} \cntLt[k] \cdot \indicator{\LCol[k] = \LCol[j]} \right)
\notag\displaybreak[0]\\
&  = \select_{\cntFt}\left( 1, \sum_{k=1}^{n} \cntL[k] \cdot \indicator{\LCol[k] < \LCol[j]} \cdot
     \indicator{k \neq i} +
     \sum_{k=1}^{j} \cntL[k]\cdot\indicator{L[k] = L[j]} \cdot
     \indicator{k \neq i} \right)
\notag\displaybreak[0]\\
&  = \select_{\cntFt}\left(1, \sum_{k=1}^{n} \cntL[k] \cdot \indicator{\LCol[k] < \LCol[j]}
     - \indicator{\LCol[j] > \LCol[i]} \quad + \right.
\notag\\
& \hphantom{ = \select_{\cntFt}\left(1, \vphantom{\sum_{k=1}^{\C[\LCol[j]]}}\right.}\left.
     \sum_{k=1}^{j} \cntL[k] \cdot \indicator{\LCol[k] = \LCol[j]}
     - \indicator{\LCol[j] = \LCol[i] \text{ and } j \geq i} \right)
\label{subeq:rankpreserve-clearing-3}\displaybreak[0]\\
& = \select_{\cntFt}\left(1, \sum_{k=1}^{n} \cntL[k] \cdot \indicator{\LCol[k] < \LCol[j]} +
    \sum_{k=1}^{j} \cntL[k]\cdot\indicator{\LCol[k]=\LCol[j]} -
    \indicator{\LF[j] \geq \LF[i]} \right)
\label{subeq:rankpreserve-clearing-4}\displaybreak[0]\\
& = \select_{\cntF}\left(1, \sum_{k=1}^{n} \cntL[k] \cdot \indicator{\LCol[k] < \LCol[j]} +
    \sum_{k=1}^{j} \cntL[k] \cdot \indicator{\LCol[k] = \LCol[j]} \right)
\label{subeq:rankpreserve-clearing-5}\displaybreak[0]\\
& = \LF_{\cntF}^{\cntL}[j] = \LF[j]
\notag
\end{align}
\end{subequations}
For the equality between \eqref{subeq:rankpreserve-clearing-3} and
\eqref{subeq:rankpreserve-clearing-4} we must ensure that 
\[ - \indicator{\LCol[j] > \LCol[i]}
   - \indicator{\LCol[j] = \LCol[i] \text{ and } j \geq i} =
   - \indicator{\LF[j] \geq \LF[i]} \]
First, if $\LCol[j] > \LCol[i]$, then certainly $\LF[j] \geq \LF[i]$ because 
$\LF[j] = \C[\LCol[j]] + \rank_\LCol(\LCol[j], j) \geq \LF[i]$ by the definition of
LF-mapping in Lemma \ref{lem:bwtinvertibility} and the precondition. In the second case,
if $\LCol[j] = \LCol[i] \text{ and } j \geq i$, we know that
\begin{align*}
\LF[j] = & \C[\LCol[j]] + \rank_\LCol(\LCol[j], j) \overset{j \geq i}{\geq} \\
         & \C[\LCol[j]] + \rank_\LCol(\LCol[i], i) \overset{\LCol[j] = \LCol[i]}{=}
           \C[\LCol[i]] + \rank_\LCol(\LCol[i], i) = \LF[i]
\end{align*}
Both cases cannot intersect, the backward way of the argumentation follows analogously.
The second equality to be explained is located between
\eqref{subeq:rankpreserve-clearing-4} and \eqref{subeq:rankpreserve-clearing-5}.
Define 
\[ \LFarg{x} \coloneqq \sum_{k=1}^{n} \cntL[k] \cdot \indicator{\LCol[k] < \LCol[x]}
   + \sum_{k=1}^{x} \cntL[k] \cdot \indicator{\LCol[k] = \LCol[x]} \]

As the difference between $\cntF$ and $\cntFt$ consists in clearing the $\LFarg{i}$-th
set bit in $\cntF$, we get following connection between $\cntF$ and $\cntFt$:
\[ \select_{\cntF}\left(1, x\right) =
   \select_{\cntFt}\left(1, x - \indicator{x \geq \LFarg{i}}\right)
   \quad \text{for }x \neq \LFarg{i}
\]

Because $\LFarg{j} \geq \LFarg{i} \Leftrightarrow 
\LF_{\cntF}^{\cntL}[j] \geq \LF_{\cntF}^{\cntL}[i] \Leftrightarrow
\LF[j] \geq \LF[i]$ holds by precondition, we can rewrite the
select connection to
\[ \select_{\cntF}\left(1, \LFarg{j}\right) =
   \select_{\cntFt}\left(1, \LFarg{j} - \indicator{\LF[j] \geq \LF[i]}\right)
   \quad \text{for }j \neq i \]
what is identical to the equality between \eqref{subeq:rankpreserve-clearing-4}
and \eqref{subeq:rankpreserve-clearing-5}.
\end{proof}
\item	Let $i$ be an integer with $\cntL[i] = \cntF[i] = 0$. Removing position $i$
	(crossed out both in $\LCol$ and $\FCol$) from $\LCol$, $\cntL$ and $\cntF$ does
	not change the mapping: Define
	\[
	\widetilde{\cntL}[j] \coloneqq \cntL[j + \indicator{j \geq i}]
	\text{~,~}
	\widetilde{\cntF}[j] \coloneqq \cntF[j + \indicator{j \geq i}]
	\text{~and~}
	\widetilde{\LCol}[j] \coloneqq \LCol[j + \indicator{j \geq i}]
	\]
	Then, for the corresponding mapping $\widetilde{\LF}^{\widetilde{\cntL}}_{\widetilde{\cntF}}$ the following holds:
	\[
	\widetilde{\LF}^{\widetilde{\cntL}}_{\widetilde{\cntF}}[j] =
	\LF_{\cntF}^{\cntL}[j + \indicator{j \geq i}] - \indicator{\LF_{\cntF}^{\cntL}[j + \indicator{j \geq i}] \geq i}
	\]
\begin{proof}
\begin{subequations}
\begin{align}
&\LFt_{\cntFt}^{\cntLt}[j]
  = \select_{\cntFt}\left(1, \sum_{k=1}^{n-1} \cntLt[k] \cdot \indicator{\widetilde{\LCol}[k] < \widetilde{\LCol}[j]} +
  \sum_{k=1}^j \cntLt[k] \cdot \indicator{\LColt[k] = \LColt[j]} \right)
\notag\displaybreak[0]\\
& = \select_{\cntFt}\left(1, \sum_{k=1}^{n-1} \cntL[k+\indicator{k \geq i}] \cdot \indicator{\widetilde{\LCol}[k] < \widetilde{\LCol}[j]} +
    \sum_{k=1}^{j} \cntL[k + \indicator{k \geq i}]\cdot\indicator{\LColt[k] = \LColt[j]} \right)
\label{subeq:rankpreserve-removal-2}\displaybreak[0]\\
& = \select_{\cntFt}\left(1, \sum_{k=1}^{n} \cntL[k] \cdot \indicator{\LCol[k] < \LCol[j+\indicator{j \geq i}]} +
    \sum_{k=1}^{j+\indicator{j \geq i}} \cntL[k] \cdot \indicator{\LCol[k] = \LCol[j+\indicator{j \geq i}]} \right)
\label{subeq:rankpreserve-removal-3}\displaybreak[0]\\
& = \LF_{\cntF}^{\cntL}[j + \indicator{j \geq i}] - \indicator{\LF_{\cntF}^{\cntL}[j + \indicator{j \geq i}] \geq i}
\label{subeq:rankpreserve-removal-4}
\end{align}
\end{subequations}
The equality between \eqref{subeq:rankpreserve-removal-2} and
\eqref{subeq:rankpreserve-removal-3} is correct because 
$\cntL[i] = \cntF[i] = 0$ holds by precondition.

For the equality between \eqref{subeq:rankpreserve-removal-3} and
\eqref{subeq:rankpreserve-removal-4}, we will reuse the notation
defined in the proof of \eqref{proof:rankpreserve:clearingvalues}, thus Line 
\eqref{subeq:rankpreserve-removal-3} is equal to 
$\select_{\cntFt}\left(1, \LFarg{j+\indicator{j \geq i}}\right)$.
As the difference between $\cntF$ and $\cntFt$ is the removal of the $i$-th
bit (which is zero), we get following connection between $\cntF$ and $\cntFt$:
\[ \select_{\cntFt}\left(1, x\right) = 
   \select_{\cntF}\left(1, x\right) - 
   \indicator{\select_{\cntFt}\left(1, x\right) \geq i}\]
Replacing $x$ with $\LFarg{j+\indicator{j \geq i}}$ and using the definition
of the generalized LF-mapping, we obtain Line \eqref{subeq:rankpreserve-removal-4}.
\end{proof}
\end{enumerate}

\clearpage

\section{Generalized LF-Mapping Computation} \label{sec:generalizedlfmappingcomputation}
The following algorithm computes the generalized LF-mapping from Lemma \ref{lem:rankpreserve}:

\begin{algorithm}[h!]
\scriptsize
\KwData{Burrows-Wheeler-Transform $\LCol$ of size $n$ over alphabet $\Sigma$ of size $\sigma$,
	bitvectors $\cntL$ and $\cntF$ of size $n+1$ with $\cntL[n+1] = \cntF[n+1] = 1$.}
\KwResult{Generalized LF-mapping $\LF^*$ of size $n$.}
\BlankLine
\DontPrintSemicolon
	let $\C$ be an array of size $\sigma$ initialized with zeros\;
	\For(\tcp*[f]{count character occurrences}){$i \gets 1$ \KwTo $n$}{
		$\C[\LCol[i]] \gets \C[\LCol[i]] + \cntL[i]$\;
	}
	\BlankLine
	$j \gets 1$\;
	\For(\tcp*[f]{compute start positions}){$i \gets 1$ \KwTo $\sigma$}{
		$cnt \gets \C[i]$\;
		$\C[i] \gets j$\;
		\While(\tcp*[f]{skip empty positions}){$cnt > 0$}{
			$cnt \gets cnt - \cntF[j]$\;
			$j \gets j + 1$\;
		}
	}
	\BlankLine
	\For{$i \gets 1$ \KwTo $n$}{
		$\LF^*[i] \gets \C[\LCol[i]]$\;
		$j \gets \C[\LCol[i]] + \cntL[i]$\;
		\While(\tcp*[f]{skip empty positions}){$\cntF[j] = 0$}{
			$j \gets j+1$
		}
		$\C[\LCol[i]] \gets j$
	}
\caption{Computation of Generalized $\LF$-mapping from Lemma
	\ref{lem:rankpreserve}\label{alg:generalizedlf}.}
\end{algorithm}

\section{Run-Block Computation} \label{sec:runblockcomputation}
The following algorithm computes the set of all run-blocks, as described in
Section \ref{sec:practicalimplementation:blockcomputation}:

\newcommand{\BE}{\mathsf{BE}}
\newcommand{\RBE}{\mathsf{RBE}}
\begin{algorithm}[h!]
\scriptsize
\SetKwFunction{runstart}{runstart}
\SetKwFunction{runheight}{runheight}
\KwData{LF-Mapping $\LF$, number of runs $R$, function $\runstart{r}$ returning
	the highest position of each run in ascending order ($\runstart{$R$} = n$)
	and function $\runheight{$r$} \coloneqq \runstart{$r+1$} - \runstart{$r$}$.}
\KwResult{the array $\RBE$ storing for each run-block $B = d-[i,j]$ either $\LF^{w_B}[i]$
	(if $B$ is width-maximal, $h_B > 1$ and $w_B > 1$) or $\LF[i]$ else.
	}
\SetKwFunction{stackpush}{\ensuremath{s}.push}
\SetKwFunction{stackpop}{\ensuremath{s}.pop}
\SetKwFunction{stacktop}{\ensuremath{s}.top}
\SetKwFunction{stackempty}{\ensuremath{s}.empty}
\BlankLine
\BlankLine
\DontPrintSemicolon
	let $\BE$ be an array of size $R$, initialized by $\BE[i] \gets \LF[\runstart{i}]$\;
	let $\RBE$ be a copy of $\BE$, and let $s$ be an empty stack\;
	\ForEach{run $r$ with $\runheight{r} > 1$} {
		$\stackpush{$r$}$\;
		\Repeat{$\stackempty{}$}{ \label{alg:run-block-computation:innerloopstart}
			$b \gets \stacktop{}$\;
			let $\widetilde{b}$ be the run with $\runstart{$\widetilde{b}$} \leq \BE[b] < \runstart{$\widetilde{b} + 1$}$\;
			\label{alg:run-block-computation:findrunbinarysearch}
			\If(\tcp*[f]{$b$ can be extended})
			   {$\BE[b]+\runheight{$b$} \leq \runstart{$\widetilde{b}+1$}$}{
				\stackpush{$\widetilde{b}$}\;
			}
			\Else(\tcp*[f]{$b$ cannot be extended}){
				\stackpop{}\;
				\If{not \stackempty{}}{
					$\BE[\stacktop{}] \gets \BE[b] + \BE[\stacktop{}] - \runstart{$b$}$ \tcp*[r]{adapt block pointer}
					\label{alg:run-block-computation:bpadapt}
					\BlankLine
					\BlankLine
					\If{$\runheight{$b$}=\runheight{$\stacktop{}$}$}{
						$\RBE[\stacktop{}] \gets \RBE[b]$ \tcp*[r]{extend run-block $\stacktop{}$}
						$\RBE[b] \gets \LF[\runstart{$b$}]$ \tcp*[r]{`remove' $b$ (not width-maximal)}
					}
				}
			}
		} \label{alg:run-block-computation:innerloopend}
	}
\caption{Computation of width-maximal run-blocks}
\label{alg:run-block-computation}
\end{algorithm}

Special attention should be given to Line \ref{alg:run-block-computation:bpadapt}.
The array $\BE$ contains
block pointers pointing to the position where a higher block on the stack was accessed.
Once this element is popped from the stack, the block pointer must be altered to the
position of the block pointer of the popped block plus the relative height offset
when accessing that block--correction is ensured by Lemma \ref{lem:lfrunparallelism}.

For analyzing the Time Complexity of Algorithm, we note that the behavior of the Loop
from Lines \ref{alg:run-block-computation:innerloopstart} to 
\ref{alg:run-block-computation:innerloopend} is equal to extracting a substring of the
original text sequence. Every time when the `extracted substring' overlaps with an
already extracted one, Line \ref{alg:run-block-computation:bpadapt} ensures that the
characters are not extracted twice ($\BE$ was modified beforehand already).
Thus, overall, the Loop is executed at most $n$ times completely using at most
additional $n$ `jumps' from Line \ref{alg:run-block-computation:bpadapt}, as each
position in the original string is extracted at most once. By implementing Line
\ref{alg:run-block-computation:findrunbinarysearch} with a simple binary search%
\footnote{Faster Methods are known for this Problem, but we want to keep it simple.}%
, the overall Time Complexity is $O(n \log r)$, although the practical runtime should
be substantially smaller.

\section{Estimations for gross-benefit and aux-tax} \label{sec:benefittaxestimators}
Let $\LCol$ be a normal $\BWT$ of with $r$ runs, and let $\widetilde{\LCol}$
and $\AUX$ be the components emerging by tunneling $\LCol$ with $t$ tunnels.
To simplify the equations, we define following additional notions:
\begin{itemize}
\item	$\rlencode$ is a function which returns the run-length-encoded
	string of $S$, i.e.\, the function replaces each maximal run of 
	character $c$ and height $h$ with the string $c\,h_k h_{k-1} \dotsm h_1$,
	where $(1\,h_k h_{k-1} \dotsm h_1)_2$ is the binary representation of $h$.
\item	$n \coloneqq \left|\rlencode( \LCol )\right|$ the size of the run-length encoded $\BWT$.
\item	$rc \coloneqq n - r$ the number of run-length symbols in $\rlencode( \LCol )$.
\item	$tc \coloneqq n - \left|\rlencode\left( \widetilde{\LCol} \right)\right|$ the number of characters
	which were removed by tunneling.
\item	$r_{h>1}$ the number of runs in $\LCol$ with height greater than 1.
\end{itemize}
Under the assumption that the two characters used for run-length-encoding occur
with same frequency, the gross-benefit (in bits) of tunneling can be estimated as follows:
\begin{align*}
\MoveEqLeft n \cdot H(\rlencode(\LCol)) - (n-tc) \cdot H(\rlencode\left(\widetilde{\LCol}\right)) \\
{}={} & 
  \underbrace{n \log n \vphantom{\log\left( \frac{rc}{2} \right)}}_{\text{full information}}
  - \underbrace{\frac{rc}{2} \log\left( \frac{rc}{2} \right)}_{\text{first run-length symbol}}
  - \underbrace{\frac{rc}{2} \log\left( \frac{rc}{2} \right)}_{\text{second run-length symbol}}
  - \underbrace{X \vphantom{\log\left( \frac{rc}{2} \right)}}_{\text{non-run-characters}}
  \\
{}-{} & \left(
  (n - tc) \log( n - tc)
  - \frac{rc-tc}{2} \log \left( \frac{rc-tc}{2} \right)
  - \frac{rc-tc}{2} \log \left( \frac{rc-tc}{2} \right)
  - X
  \right) \\
{}={} &
  n \log n
  - rc \log rc
  + rc \\
{}-{} &
  n \log(n - tc)
  + rc \log(rc-tc)
  - rc
  + tc \log(n - tc)
  - tc \log(rc - tc)
  + tc \\
{}={} &
  n \log\left( \frac{n}{n-tc} \right)
  - rc \log\left( \frac{rc}{rc-tc} \right)
  + tc \left( 1 + \log\left( \frac{n-tc}{rc-tc} \right) \right)
\end{align*}
To estimate the tax required to save the encoding of $\AUX$, we remind that
$\AUX$ is a string of length $r_{h>1}$ containing $t$ symbols for the start of the tunnels,
$t$ symbols for the end of the tunnels as well as $r_{h>1}-2t$ ``zero-symbols''.
Assuming that the tunnel indicator symbols are evenly distributed over the full length
of $\AUX$, $\rlencode(\AUX)$ contains approximately $2\cdot t + 2\cdot t\cdot h$ symbols,
where $h \coloneqq \log\left( \frac{r_{h>1} - 2t}{2t} \right)$ is the average logarithmic height of a run.
Consequently, the tax (in bits) for the encoding of $\AUX$ is given by
\begin{align*}
\MoveEqLeft 2\cdot t \cdot (h+1) \cdot H(\rlencode(\AUX)) \\
{}={} & 
  \underbrace{2 \cdot t \cdot (h+1) \cdot \log( 2 \cdot t \cdot (h+1) )}_{\text{full information}}
  - \underbrace{t \log t}_{\text{tunnel starts}}
  - \underbrace{t \log t}_{\text{tunnel ends}}
  \\
{}-{} &
  \underbrace{2 \cdot t \cdot \log( 2 \cdot t )}_{\text{``zero-symbols''}}
  - \underbrace{t \cdot (h-1) \cdot \log( t \cdot (h-1) )}_{\text{first run-length symbol}}
  - \underbrace{t \cdot (h-1) \cdot \log( t \cdot (h-1) )}_{\text{second run-length symbol}}
  \\
{}={} &
  2t \log( 2 \cdot t \cdot (h+1) )
  - 2t \log t
  - 2t \log( 2 \cdot t ) 
  + 2t \log( 2\cdot t \cdot (h-1) )
  \\
{}+{} &
  2 \cdot t \cdot h \cdot \log( 2\cdot t \cdot (h+1) )
  - 2 \cdot t \cdot h \cdot \log( 2\cdot t \cdot (h-1) )
  \\
{}={} &
  2t\left( 1 + \log( h^2 - 1 ) \right)
  + 2\cdot t \cdot h \cdot \log\left( 1 + \frac{2}{h-1} \right)
\end{align*}

\section{Worst-Case Time of Algorithm \ref{alg:blockchoice}} \label{sec:algblockchoicetime}

This section analyzes the worst-case-time of Algorithm \ref{alg:blockchoice}.
Beforehand however, we introduce a lemma showing some properties of run-blocks.

\begin{lemma} \label{lem:runblockprops}
Let $\RB$ be a set of width-maximal run-blocks of a $\BWT$ of length $n$.
\begin{enumerate}
\item	\label{lem:runblockprops:collisioncount}
	The number of pairwise collisions in $\RB$ is limited by $\sum_{B \in \RB} w_B$.
\item	\label{lem:runblockprops:widthsum}
	The sum of all widths $\sum_{B \in \RB} w_B$ is limited by $n$.
\end{enumerate}

\begin{proof}~
\begin{enumerate}
\item	Since $\RB$ is a set of width-maximal run-blocks, it contains only compensable collisions,
	so every collision between two blocks consists of an outer and inner block. To bound the
	collision count, it is enough to count all inner collisions. As each block
	of $\RB$ is also height-maximal, a block $B \in \RB$ can have at most one collision per
	column, and thus at most $w_B$ collisions, what overall yields a bound of at most
	$\sum_{B \in \RB} w_B$ collisions.
\item	To see that $\sum_{B\in \RB} w_B$
	is bounded by $n$, we note that each block $B \in \RB$ has at least one position in its
	start interval which is unshared--otherwise, $B$ would not be width-maximal.
	Thus, the rows starting at those positions must be non-overlapping strings in the original text,
	so their width, $\sum_{B\in \mathcal B} w_B$ must be bounded by $n$. 
\end{enumerate}~\\[-2\baselineskip]
\end{proof}
\end{lemma}

\begin{theorem} \label{thm:algblockchoicetime}
Algorithm \ref{alg:blockchoice} applied to a set $\RB$ of run-blocks originating
from a $\BWT$ of size $n$ requires $O(n \log |\RB|)$ time, given that the function
$\FuncSty{score}$ and the benefit- and tax-estimators can be computed and updated in $O(1)$.

\begin{proof}
A collision graph as described in Section \ref{sec:practicalimplementation:blockchoice} can be computed
in $O(\sum_{B\in \RB} w_B)$ time using block information and the LF-mapping. As
the graph allows to visit all colliding blocks of a block in optimal time, and
a node removal can be performed in time proportional to the degree of a node,
Algorithm \ref{alg:blockchoice} is dominated by the time for updating colliding
blocks. Each update requires $O( \log|\RB| )$ time using a heap, with 
Lemma \ref{lem:runblockprops} we get an overall time of $O(n \log |\RB| )$.
\end{proof}
\end{theorem}

It should be mentioned that Theorem \ref{thm:algblockchoicetime} can be applied
to both a $\BWT$ with and without run-length-encoding, and that the initial
score computation for run-length-encoding can be performed in $O(n)$ using the $\LF$-mapping
in conjunction with Lemma \ref{lem:runblockprops}.

\end{document}